\documentstyle[12pt]{article}
\begin{document}
\baselineskip 3.9ex
\def\be{\begin{equation}}
\def\ee{\end{equation}}
\def\ba{\begin{array}{l}}
\def\ea{\end{array}}
\def\bea{\begin{eqnarray}}
\def\eea{\end{eqnarray}}
\def\no#1{{\tt  hep-th#1}}
\def\nn{\nonumber}
\def\nl{\hfill\break}
\def\ni{\noindent}
\def\bibi{\bibitem}
\def\c#1{{\hat{#1}}}
\def\eq#1{(\ref{#1})}
\def\pgap{\vspace{1.5ex}}
\def\ggap{\vspace{10ex}}
\def\gap{\vspace{3ex}}
\def\del{\partial}
\def\o{{\cal O}}
\def\z{{\vec z}}
\def\re#1{{\bf #1}}
\def\av#1{{\langle  #1 \rangle}}
\def\S{{\cal S}}
\def\sbh{S_{\rm BH}}
\renewcommand\arraystretch{1.5}

\begin{flushright}
TIFR-TH-00/31\\
\end{flushright}
\begin{center}
\vspace{2 ex}
{\large{\bf Lectures on the Microscopic Modeling 
of the 5-dim. Black Hole of IIB String Theory and the 
$D_1/D_5$ System }}$^*$\\
\vspace{3 ex}
Spenta R. Wadia\\
{\sl Department of Theoretical Physics ,}\\
{\sl Tata Institute of Fundamental Research,}\\
{\sl Homi Bhabha Road, Mumbai 400 005, INDIA. }\\
{\sl email: wadia@tifr.res.in}\\
\vspace{10 ex}
\pretolerance=1000000
\bf ABSTRACT\\
\end{center}
\vspace{1 ex}
In these notes we review the theory of the microscopic modeling of the 5-dim. 
black hole of type IIB string theory in terms of the $D1-D5$ brane system.
The emphasis here is more on the brane dynamics rather than on supergravity solutions.
We present a discussion of the low energy brane dynamics and account for black hole thermodynamics and Hawking radiation rates. These considerations are valid in the regime of supergravity due to the non-renormalization of the low energy dynamics in this model.
\vfill
\hrule
\vspace{0.5 ex}
{\small $^*$ Based on lectures given at the, ``Advanced School on Supersymmetry in the
Theories of Fields, Strings and Branes'' , 
Santiago de Compostela, Spain, July 1999; 
``Workshop on String Theory'', at the Mehta Research Institute, 
Allahabad, India, Oct. 1999; 
Isfahan String School and Workshop, 
Iran, May 2000.}

\clearpage

\newpage

\section{\bf Quantum Mechanics and General Relativity} 
The application of quantum field theory to general relativity (GR)
leads to some basic problems:

\begin{enumerate}
\item The problem of ultra-violet divergences renders GR an ill-defined
quantum theory. This specifically means that if we perform a
perturbation expansion around flat Minkowski space-time (our world!)
then to subtract infinities from the divergent diagrams we have to add
an infinite number of terms to the Einstein-Hilbert action with
coefficients that are proportional to appropriate powers of the
ultraviolet cutoff.

There is good reason to believe that string theory solves this
ultra-violet problem because the extended nature of string
interactions have an inherent ultra-violet cutoff given by the
fundamental string length $\sqrt{\alpha'}$. One also knows that in
string theory the Einstein-Hilbert action emerges as a low energy
effective action for energy scales much larger than the string length
and Newton's constant (in 10-dim.) is given by,
\be
G_{10}=\kappa _{10}^2= 8 \pi^6 g_s^2 \alpha'^4
\ee
where $g_s$ is the string coupling. 

\item The solutions of general relativity can be singular. There are a
variety of singularities that have been encountered. Examples are i)
the singularity of the black hole solution, ii) the singularities
encountered in various brane solutions of supergravity, iii)
singularities of the cosmological solutions of GR etc. A quantum
theory of gravity must present an understanding about what are good
and bad singularities in the sense whether one can have a well defined
quantum mechanics in their presence. String theory has `resolved' some
of these singularities, but a complete understanding of the issue of
singularities is still lacking.
 
\item While the above problem is related to the high energy (short
distance) behavior of GR, there exists another problem when we
quantize matter fields in the presence of a black hole which does not
involve high energy processes.  This problem is called the information
puzzle and in the following we shall explain the issue and also
summarize the attempts within string theory to resolve the puzzle in a
certain class of black holes.

String theory has been proposed as a theory that describes all
elementary particles and their interactions. Presently the theory is
not in the stage of development where it can provide quantitative
predictions in particle physics. However in case this framework
resolves some logical problems that arise in the applications of
quantum field theory to general relativity, then it is a step forward for
string theory. 

\item Finally there is the problem of the cosmological constant, which is getting 
renewed attention in recent times.

In these notes we will discuss only point 3. We will focus on
the black hole solution of IIB string theory and discuss it's
modeling by the $D_1-D_5$ system of branes. We will describe the low
energy excitations of this system and learn how they couple to the
bulk supergravity degrees of freedom using Maldacena duality. We will
present the calculation of the Hawking rate for a class of massless
particles which agrees with supergravity calculations due to the high
degree of supersymmetry of the $D_1-D_5$ system.
\end{enumerate}

\par

\section{\bf Organization of the Notes}

\begin{itemize}
\item  Sections 3 and 4 present a general description of black hole
thermodynamics and the information puzzle.
\item Section 5 presents the string theory framework for black holes.
\item Section 6 presents various supergravity solutions of
relevance to our discussion: The BPS and the non-BPS black hole
solution, the Maldacena limit and $AdS_3\times S^3$, and the solution
with a non-zero value of the Neveu-Schwarz B-field in the $4$ compact
dimensions. We also discuss the semi-classical derivation of Hawking
radiation.
\item Section 8 presents the $D_1-D_5$ system and the $N=4$, $U(Q_1)\times U(Q_5)$ gauge theory in 2-dimensions. We discuss
low energy degrees of freedom and the conformally invariant sigma model
at the infrared fixed point.
\item Section 9 presents the discussion of $D_1$ branes as
solitonic strings of the $D_5$ gauge theory. We discuss the moduli space of
instantons which forms the target space of the solitonic strings.
\item Sections 10-12 discuss the ${\cal N}=4$ super conformal algebra,
the ${\cal N}=(4,4)$ SCFT on the orbifold $(\tilde{T}^4)^{Q_1
Q_5}/S(Q_1 Q_5) $, and the classification of states of the SCFT in
terms of the supergroup $SU(1,1|2)$. We identify the maximally twisted sector of the SCFT with the set of states that constitute the black hole.
\item Section 13 compares the near horizon supergravity moduli and 
correspondence with SCFT operators.
\item Section 14 discusses the microscopic derivation of Hawking
radiation.
\item Section 15 discusses some future directions.
\end{itemize}

In sections 10-12 there is overlap with Justin's thesis \cite
{Dav-thesis}.  The details of the construction of all the chiral
primaries has been presented there.

\section{\bf The Classical Black Hole Horizon}

Classically a black hole is a solution of the GR equations, and it is
characterized by an event horizon, which is a null surface. The
horizon is a one way gate, in the sense that once we are inside it we
cannot get out because of the causal structure of the black hole
spacetime. Physically one can imagine the formation of an event
horizon due to the bending property of light by the matter that makes
up the black hole.

Let us list a few properties of classical black holes: (see e.g. the text book by Wald \cite{wald})

Firstly the event horizon has an area and there is a area law which
states that in any adiabatic process involving black holes the final area of the
event horizons is never less than the initial area(s):
\be
\label{one}
A_{12} \geq A_1 + A_2 
\ee
The `no hair theorems', tell us that the state of a classical black
hole is completely characterized by its mass, angular momentum and
global gauge charges. In particular the area of the event horizon
depends only on these quantities. If we perturb a black hole then the
perturbation decays in Planck time, and the new state of the black
hole is again characterized by a event horizon whose area has
increased and is characterized by the changed mass, angular
momentum or charge of the final state. 

The area law (\ref{one}) prompted Bekenstein \cite{bek} to associate a entropy with the black
hole that is proportional to its area: 
\be
\label{two}
S=aA
\ee
where `a' is a universal constant.  The area law (\ref{one}) then
resembles the second law of thermodynamics where the black hole is
treated as a macroscopic object,
\be
\label{three}
S_{12} \geq S_1 + S_2 
\ee
>From the viewpoint of classical general relativity there is no
information puzzle because the stuff that went inside a black hole
stays inside because the horizon is a one way gate.

\section{\bf Quantum Mechanics and The Information Puzzle of Black Holes}

In the quantum theory since the absorption process is described by the
matrix element of a hermitian hamiltonian, the emission amplitude is
necessarily non-zero. Black holes radiate. 

Application of quantum field theory to matter propagating in a black
hole background leads to the following results which we briefly summarize:

Black holes behave like black bodies. They emit thermal
radiation and they are characterized by a temperature which
depends only on the mass, angular momentum and the global charges of
the black hole. The fundamental formula for the temperature, due to
Hawking \cite{Hawk} is given by,
\be
\label{temp}
T = \frac{\hbar \kappa}{2 \pi}
\ee
$\kappa$ is surface gravity (acceleration due to gravity
felt by a static observer) at the horizon of the black hole.
For a Schwarzschild black hole: 
\be 
\kappa = \frac{1}{4 G_N M} 
\ee
The constant of proportionality is determined using the first law of
thermodynamics and the temperature formula: $TdS=dM$,

\be
\label{HB}
\S_{\it bh} = a A_h, \qquad a= \frac{c^3}{4 G_N \hbar}
\ee
Using this we can now interpret (\ref{three}) as the 
second law of black hole thermodynamics.

Formula (\ref{HB}), called the Bekenstein-Hawking formula is very
fundamental because it is a formula that counts the effective
degrees of freedom of the black-hole. $a^{-1}$ is a basic unit of area and 
it has all the 3 fundamental constants in it.

The Hawking radiation as calculated in semi-classical GR is a mixed
state.  It turns out to be difficult to calculate the correlations
between the ingoing and outgoing Hawking particles in the standard
framework of general relativity. Such a calculation would require a
good quantum theory of gravity where controlled approximations are
possible.

If we accept the semi-classical result that black holes emit radiation that is EXACTLY thermal then it leads
to the information puzzle:

Initially the matter that formed the black hole is in a pure quantum
mechanical state. Here in principle we know all the quantum mechanical
correlations between the degrees of freedom of the system.  In case
the black hole evaporates completely then the
final state of the system is purely thermal and hence it is a mixed
state. This evolution of a pure state to a mixed state is in conflict
with the standard laws of quantum mechanics which involve unitary time
evolution of pure states into pure states.

Hence we either have to modify quantum mechanics, as was advocated by
Hawking \cite{Hawkbook}, or as we shall argue, the other possibility is to replace the
paradigm of quantum field theory by that of string theory. In string
theory we retain quantum mechanics and resolve the information puzzle
(for a certain class of black holes) by discovering the microscopic
degrees of freedom of the black hole. In string theory the Hawking
radiation is NOT thermal and in principle we can reconstruct the
initial state of the system from the final state. 

In standard statistical mechanics, for a system with a large number of
degrees of freedom we introduce a density matrix to derive the
thermodynamic description. The same thing can be done for black holes
in string theory. In this way the thermodynamic formulas for black
hole entropy and decay rates of Hawking radiation can be derived from
string theory. In particular the Bekenstein-Hawking formula is derived
from Boltzmann's law:
\be
\label{Boltz}
S=\ln \Omega
\ee
where $\Omega$ is the number of micro-states of the system.

It is well worth pointing out that the existence of black holes in
nature (for which there is mounting evidence) compels us to resolve
the conundrums that black holes present. One may take recourse to the
fact that for a black hole whose mass is a few solar masses the Hawking
temperature is very tiny ($\sim 10^{-8}$ degs. Kelvin), and not of any
observable consequence. However the logical problem that we have
described above cannot be wished away and its resolution makes a
definitive case for the string paradigm as a correct framework for
fundamental physics as opposed to standard local quantum field theory.

\section{\bf The String Theory Framework for Black Holes}

The basic point in the string theory description is that a black
hole is described by a density matrix:
\bea
\label{density}
\rho & = & {1\over \Omega }\sum_{i}|i\rangle \langle i|  \nn \\
S & = & \ln{ \Omega}
\eea
where $|i\rangle $ is a micro-state. 

Given this we can calculate formulas of black hole thermodynamics just
like we calculate the thermodynamic properties of macroscopic objects
using standard methods of statistical mechanics. Here the quantum
correlations that existed in the initial state of the system are in
principle all present and are only erased by our procedure of defining the
black hole state in terms of a density matrix. In this way one can
account for not only the entropy of the system which is a counting
problem but also the rate of Hawking radiation which depends on
interactions.

Let us recall the treatment of radiation coming from a star, or a lump
of hot coal. The `thermal' description of the radiation coming is the
result of averaging over a large number of quantum states of the coal.
In principle by making detailed measurements on the wave function of
the emitted radiation we can infer the precise quantum state of the
emitting body. For black holes the reasoning is similar.

Hence in the string theory formulation the black hole can exist as a
pure state: one among the highly degenerate set of states that are
characterized by a small number of parameters. Let us also note that
in Hawking's semi-classical analysis, which uses quantum field theory
in a given black-hole space-time, there is no possibility of a
microscopic construction of the black hole wave functions.

We summarize the 4 basic ingredients we need in string theory to
calculate Hawking radiation from low temperature near extremal black
holes:
\begin{enumerate}
\item The microscopic constituents of the black hole. In the case of the 
5-dim. black hole of type IIB string theory the microscopic modeling
is in terms of a system of $D_1-D_5$ branes wrapped on $S^{1}\times M_{4}$,
where $M_{4}$ is a 4-dim. compact manifold, which can be either $T^{4}$
or $K_{3}$. Here we will consider $T^{4}$.
\item The spectrum of the low energy degrees of freedom of the bound state of 
the $D_1-D_5$ system. Usually these are arrived at weak coupling
and we need to know if the spectrum survives at strong coupling.
\item The coupling of the low energy degrees of freedom to supergravity modes.
\item The description of the black hole as a density matrix. This 
implies expressions for decay and absorption probabilities which are related to
S-matrix elements between initial and final states of the black hole.
\end{enumerate}

The decay probability from a state 
$|i>$ to a state $|f>$ is given by
\be
\label{probdecay}
P_{\it decay}(i\rightarrow f)=\sum_{\it i,f}\frac{1}{\Omega _{\it f}}|<f|S|i>|^{2}
\ee 
The absorption probability from a state
$|i>$ to a state $|f>$ is given by
\be
\label{probabs}
P_{\it abs}(i\rightarrow f)=\sum_{\it i,f}\frac{1}{\Omega _{\it i}}|<f|S|i>|^{2} 
\ee 
In the above formulae $ \Omega _{\it f}$ and $\Omega _{\it i}$ refer to the number of 
final and initial states respectively.

One of the important issues in this subject is that 1. and 2. are
usually known in the the case when the effective open string coupling
is small. In this case the Schwarzschild radius $R_{sch}$ of the
black hole is smaller than the string length $l_s$ and we have a
complicated string state.  As the coupling is scaled up we go over to
the supergravity description where $R_{sch} \gg l_s$ and we have a
black hole. Now it is an issue of dynamics whether the spectrum of the
theory undergoes a drastic enough change, so that the description of
states in weak coupling which enabled a thermodynamic description is
still valid. In the model we will consider we will see that the
description of the weak coupling effective lagrangian goes over to strong coupling
because of supersymmetry. It is an outstanding challenge to understand
this problem when the weak coupling theory has little or no
supersymmetry \cite{susskind,polhoro,venezia}

\section{\bf Black Holes of IIB String Theory: \\ 
Supergravity Solutions}
We will now present a summary of the SUGRA solutions of relevance to
the $D_1-D_5$ system.  This will include the BPS and near BPS black
hole solutions, and the near horizon geometry of the D1-D5 system. We also
discuss the geometry in the presence of the vev of the Neveu-Schwarz
$B_{NS}$ with components along the directions of the internal space
$T^4$. There is a huge literature on this subject and we refer the reader to the review by Mandal \cite{gugireview}. The material relevant to our discussion can be found is \cite {Str-Vaf96,Cal-Mal96,Mal96, Hor-Mal-Str96,Obe-Pio98,DMWY}.

\subsection{The BPS Black Hole}
Let us begin by describing type IIB string theory in 10-dimensions.
This string theory has 32 real sypersymmetries. It's massless bosonic content 
in the NS-NS sector consists of the metric $G_{a,b}$, the dilaton $\phi $ and the 
2-form $B^{(2)}_{NS}$. The R-R sector consists of the gauge potentials $C^{n}$, 
$n=0,2,4$. The low energy effective action is given by 
\bea
\label{action}
S_{IIB}&=&{1\over 2\kappa^2} \int d^{10} x \sqrt{-G} \bigg\{
e^{-2\phi}
\left(R + 4 (\nabla \phi)^2 - {1\over 2.3 !} (H^{(3)})^2\right) -
{1\over 2.3 !} (F^{(3)})^2   \\  \nonumber 
&+& - {1\over 4.5 !} (F^{(5)})^2\bigg\} + {1\over 4\kappa^2} \int C^{(4)}
\wedge F^{(3)} \wedge H^{(3)} ,
\eea
where $(H^{(3)})^2 = H^{(3)}_{\rm MNP} H^{(3){\rm MNP}} , \ (F^{(n)})^2
=
F^{(n)}_{M_1 \cdots M_n} F^{(n)M_1 \cdots M_n}$ and, using the standard
form notation,
\be
\label{4}
H^{(3)} = dB^{(2)}_{NS} , \ \ F^{(3)} = dC^{(2)} , \ \ F^{(5)} =
dC^{(4)} - {1\over 2} C^{(2)} \wedge H^{(3)} + {1\over 2} B^{(2)} \wedge
F^{(3)}.
\ee
The self-duality constraint, $\ast F^{(5)} = F^{(5)}$, is imposed at
the level of the equations of motion. Also, $\kappa^2 = 8 \pi G_{10}$,
where
$G_{10} = 8\pi^6 g_s^2 \alpha^{\prime 4}$ is the 10-dimensional Newton's
constant (in the convention that the dilaton, $\phi$, vanishes
asymptotically).

Let us now present the supergravity solution that preserves 4 out of the 32 SUSYs
of the original theory. A simple ansatz is to consider all the bosonic fields in
(\ref{action}) to be zero except the metric $G_{a,b}$, the dilaton $\phi $ and the 
Ramond 2-form $C^{2}$. We compactify the $6,7,8,9$ directions on a
torus $T^4$ of volume $V_4$ and the $x_5$ direction on a circle of
radius $R_5$. We then wrap $Q_5$ D5-branes along the directions
$5,6,7,8,9$ and $Q_1$ D1-branes along the $x_5$ direction. We
introduce $N$ units of momentum along the $x_5$ direction in order to
obtain a black hole of finite horizon area. The supergravity solution
with these boundary conditions is given by:
\bea
\label{d1_d5}
ds^2 &=& f_1^{-\frac{1}{2}} f_5^{-\frac{1}{2}} (-dt^2 + dx_5^2
+ k(dt - dx_5)^2 ) 
+ f_1^{\frac{1}{2}} f_5^{\frac{1}{2}} (dx_1^2 + \cdots + dx_4^2) 
\\ \nonumber 
    & & + f_1^{\frac{1}{2}} f_5^{-\frac{1}{2}} 
(dx_6^2 + \cdots + dx_9^2),
\\ \nonumber
e^{-2 \phi} &=&\frac{1}{g_s^2} f_5 f_1^{-1} , \\ \nonumber
C^{(2)}_{05} &=& \frac{1}{2} (f_1^{-1} -1), \\ \nonumber
F^{(3)}_{abc} &=& (dC^{(2)})_{abc}
=\frac{1}{2}\epsilon_{abcd}\partial_{d} f_5, \;\;\;\;
 a, b, c, d = 1, 2, 3, 4 
\eea
where $f_1$, $f_5$ and $k$ are given by
\be
f_1 = 1 + \frac{16 \pi ^4 g_s \alpha '^3 Q_1}{V_4 r^2}, \;
f_5= 1+ \frac{ g_s \alpha' Q_5}{r^2},\;
k= \frac{16 \pi ^4 g_s ^2 \alpha '^3 N}{V_4 R_5^2 r^2}
\ee
Here $r^2 = x_1^2 + x_2^2 + x_3^2 + x_4^2$ denotes the distance
measured in the transverse direction to all the D-branes. 

The horizon in the above solution occurs at $r=0$ and we can read off the horizon area and hence the entropy which is given by,
\be
\label{chp1:entropy}
S= 2\pi \sqrt{Q_1 Q_5 N}
\ee
The mass of the black hole for the above solution turns out to be
a linear combination of it's 3 charges,
\be
\label{mass-bps}
M=\frac{1}{g_s^2} (a_1g_sQ_1 + a_2g_sQ_5 + a_3g_s^2N)
\ee
where 
\be
\ba
\nonumber
a_{1}=\frac{R}{\alpha'} \nonumber \\
a_{2}=\frac{RV_4}{16\pi^4\alpha'^3} \nonumber \\
a_{3}=\frac{1}{R}
\ea
\ee
Anticipating the microscopic modeling, this means that we have a marginal bound state with zero binding energy.
Also, since this is an extremal black hole it's Hawking temperature is zero, 
a fact which will have an obvious explanation in the microscopic theory.

\subsection{ The Near Extremal Limit and non-zero Hawking Temperature}  
In order to have a black hole with a non-zero temperature we have to consider a 
non-BPS and non-extremal black hole solution. This solution preserves none of the original 
32 supersymmetries of the type IIB theory and can be obtained 
by
allowing the total momentum N to be distributed in both directions
around the $x_{5}$ direction. The solution in 10-dims. is given by:
\bea
\label{chp1:nonextremal}
e^{-2 \phi} &=& \frac{1}{g_s^2}
\left(1 + \frac{r_5^2 }{r^2} \right) 
\left( 1 + \frac{r_1^2}{r^2} \right)^{-1} 
,\\  \nonumber
F^{(3)} &=& \frac{2r_5^2}{g_s}\epsilon_3 +
2g_se^{-2\phi}r_1^2 *_6\epsilon_3 , \\ \nonumber
ds^2 &=& 
\left( 1 + \frac{r_1^2}{r^2} \right)^{-1/2} 
\left(1 + \frac{r_5^2}{r^2} \right) ^{-1/2} 
\left[ -dt^2 + dx_5^2 \right.  \\  \nonumber
&+& \frac{r_0^2}{r^2} (\cosh\sigma dt + \sinh\sigma dx_5)^2 +
\left.
\left( 1 + \frac{r_1^2}{r^2} \right) 
g_sQ_5(dx_6^2 + \ldots dx_9^2) \right] \\ \nonumber
&+&\left( 1 + \frac{r_1^2}{r^2} \right)^{1/2} 
\left(1 + \frac{r_5^2}{r^2} \right) ^{1/2} 
\left[ \left( 1- \frac{r_0^2}{r^2} \right)^{-1} dr^2 + r^2 d\Omega_3^2
\right],
\eea
where $*_6$ is the Hodge dual in the six dimensions $x_0, \ldots ,x_5$
and $\epsilon_3$ is the volume form on the unit three-sphere. $x_5$ is
periodically identified with period $2\pi R_5$ and directions $x_6,
\ldots ,x_9$ are compactified on a torus $T^4$ of volume $V_4$.
$\Omega_3$ is the volume of the unit three-sphere in the transverse
directions. This solution 
is parameterized by six independent quantities:
$r_1, r_5, r_0, \sigma,  R_5$ and $V_4$. These are related to the 
number of D1-branes , D5-branes and Kaluza-Klein momentum on $x_5$  
as follows,
\bea
\label{fields}
Q_1 &=&\frac{V_4}{64\pi^6 g_s^2\alpha^{\prime 3}}\int e^{2\phi}*_6 F^{(3)} 
= \frac{V_4 r_1^2}{16\pi^4\alpha^{\prime 3} g_s} ,
\\ \nonumber
Q_5 &=&\frac{1}{4\pi^2 \alpha^{\prime}} \int F^{(3)}  =
\frac{r_5^2}{g_s\alpha^{\prime}} , \\ \nonumber
N &=& \frac{R_5^2 V_4 r_0^2}{32\pi^4\alpha^{\prime 4} g_s^2} \sinh
2\sigma .
\eea
$r_o$ is the non-extremality parameter. At $r_o=0$, the two classical horizons coincide.
On compactifying this solution to five dimensions using the
Kaluza-Klein ansatz one obtains a five-dimensional black hole with a
horizon at $r=r_0$. 
 The entropy and the mass of this black hole is given
by
\bea
\label{mass_entropy}
S&=& \frac{A}{4G_5}= \frac{2\pi^2 r_1 r_5 r_0 \cosh 2\sigma}{4G_5}, \\
\nonumber 
M&=& \frac{\pi}{4 G_5}(r_1^2 + r_5^2 +\frac{r_0^2\cosh 2 \sigma}{2} ),
\eea
where the five-dimensional Newton's constant is 
\be
G_5= \frac{4\pi^5\alpha^{\prime 4} g_s^2}{V_4 R_5}
\ee

Let us now discuss the restrictions on the various parameters which
result from the requirement that the above solution makes
sense in the quantum theory and that we are actually describing a macroscopic
black hole whose horizon is much larger than the string length 
$l_{s}=\sqrt{\alpha ^{'}}$. The above classical solution has a quantum significance
only if the string coupling $g_{s}\rightarrow 0$. 
This implies that the Newton coupling $ G_5\rightarrow 0$, and hence the 
entropy formula 
(\ref{chp1:entropy}) implies that we have a finite horizon area only if 
\bea
\label{class_limit}
g_s\rightarrow 0, \\  \nonumber
\mbox{with} \; g_sQ_1, \; g_sQ_5, \; g_s^2N \; \mbox{fixed}.
\eea
The formulae in \eq{fields} indicate that this is also equivalent to
\bea
g_s\rightarrow 0, \\ \nonumber
\mbox{with} \; r_1,\;  r_5, \; r_n \; \mbox{fixed}.
\eea
where $r_N=r_0 \sinh \sigma$.
For a macroscopic black hole we require that the string length is much
smaller than the horizon area, or equivalently from
(\ref{mass_entropy}) we conclude that $r_1, r_5, r_N \gg l_{s}$. This implies
\be
\label{l_1}
g_sQ_1>>1, \; g_sQ_5>>1, \; g_s^2N >>1
\ee
Since $g_sQ_1$, $ g_sQ_5$ correspond to the effective open string coupling constants,
the macroscopic black hole exists at strong coupling! 

The non-BPS black hole has a small Hawking temperature given by
\be
T_H = \frac{r_o}{2\pi r_1r_5\cosh {\sigma}}
\sim \frac{r_o\exp -\sigma}{\pi r_1r_5} \ll 1
\ee
In the near extremal limit, when for large $\sigma$, $r_o\sim \exp
-\sigma$, we see that $T_H \sim o(r_o^2)$.

We also note that the black hole has a positive specific heat
$\Delta M = cT_H^2 > 0$.
This is unlike the case of the Schwarzschild bh : $\Delta M < 0$.

\subsection{The Near Horizon Limit of Maldacena}
In this section we will exhibit the form of the classical solution in the so called 
near horizon limit of Maldacena\cite{malda-dual}. 
To explain the basic point let us study the metric of the black hole with the KK charge 
$N=0$. In this case the horizon area shrinks to zero, but that is not relevant to the 
physical point we want to make. The metric then takes the form,
\bea
\label{d1_d5_nohor}
ds^2 &=& f_1^{-\frac{1}{2}} f_5^{-\frac{1}{2}} (-dt^2 + dx_5^2)
+ f_1^{\frac{1}{2}} f_5^{\frac{1}{2}} (dx_1^2 + \cdots + dx_4^2)
\\ \nonumber
& & + f_1^{\frac{1}{2}} f_5^{-\frac{1}{2}}
(dx_6^2 + \cdots + dx_9^2),
\\ \nonumber
e^{-2 \phi} &=& \frac{1}{g_s^2} f_5 f_1^{-1} , \\ \nonumber
C^{2}_{05} &=& \frac{1}{2} (f_1^{-1} -1), \\ \nonumber
F^{(3)}_{abc} &=& (dC^{(2)})_{abc}
=\frac{1}{2}\epsilon_{abcd}\partial_{d} f_5, \;\;\;\;
a, b, c, d = 1, 2, 3, 4
\eea
where $f_1$ and $f_5$ are given by
\be
f_1 = \frac{16 \pi ^4 g_s \alpha '^3 Q_1}{V_4 r^2}, \;\;\;\;
f_5= \frac{ g_s \alpha' Q_5 }{r^2},\;
\ee
Here $r^2 = x_1^2 + x_2^2 + x_3^2 + x_4^2$ denotes the 
distance measured in the transverse direction to all the D-branes.

The basic idea of the near horizon limit is that, near the horizon of a black 
hole, the energies of particles as seen by the asymptotic observer get red-shifted:
\be
E_{\infty}=\sqrt{G^{00}}E
\ee
In the metric at hand the red-shit factor is
\be
\sqrt G^{00} = (f_1f_5)^{-1/4}
\ee
Clearly as $r\rightarrow \infty$ the red shift factor is unity. However near the 
horizon we get the equation
\be
E_{\infty} = \frac{r}{R}E
\ee
where $R^2 \sim \alpha '\sqrt{ g_s^2Q_1Q_5}$ is the typical length
scale that characterizes the geometry. For $r \ll R$ we see that the
energy observed by the asymptotic observer goes to zero for finite
values of E. This means that near the horizon (characterized by large
R) an excitation of arbitrary energy looks massless.  For massless
modes this means that they have almost infinitely long wavelengths and
for massive modes they appear as long wavelength massless excitations.
If one examines the potential energy of a particle in the above geometry then
in the near horizon limit the potential barrier becomes very high so
that the modes near the horizon cannot get out. In the exact limit 
of $Q_1$ and $Q_5$ going to infinity the horizon degrees of freedom
become exactly massless and decouple from the bulk degrees of freedom.
As we shall see later it is in this limit that the bulk string theory
is dual to a SCFT which also exhibits massless behavior in the
infrared.

A more precise scaling limit of the geometry is given by
\bea
\alpha' \rightarrow 0, \;&\;&\; \frac{r}{\alpha '} 
\equiv U=  \mbox{fixed} \\  \nonumber
v \equiv \frac{V_4}{16\pi^4\alpha^{\prime 2}} = \mbox{fixed}, \;&\;&\;
g_6 = \frac{g_s}{\sqrt{v}} = \mbox{fixed}
\eea
In this limit the metric in (\ref{d1_d5_nohor}) becomes
\be
\label{near_horizon}
ds^2 = \alpha ' \left[ \frac{U^2}{g_6\sqrt{Q_1Q_5}} (-dx_0^2 + dx_5^2)
+ g_6\sqrt{Q_1 Q_5} \frac{dU^2}{U^2} + g_6\sqrt{Q_1Q_5} d\Omega_3^2
\right] + \sqrt{\frac{Q_1}{vQ_5}}(dx_6^2 + \ldots dx_9^2)
\ee
Thus the near horizon geometry is that of $AdS_3\times S^3\times T^4$.
Our notation for coordinates here is as follows: $AdS_3:
(x_0, x_5, r);\, S^3:(\chi, \theta, \phi);\, T^4:(x_6, x_7, x_8,
x_9)$. $r, \chi,\theta,\phi$ are spherical polar coordinates for the
directions $x_1,x_2,x_3,x_4$. 
The radius of $S^3$ 
and the anti-de Sitter space 
is $R = \sqrt{\alpha'} (g_6^2 Q_1Q_5)^{1/4}$.

Note that the effective string coupling in the near horizon limit is given by
\be
g_{eff}=g_6\sqrt {Q_1/Q_5}
\ee

The formulas for the blackhole entropy and temperature, which depend
only on the near horizon properties of the geometry, do not change in
the near horizon limit.

It is important to mention the symmetries of the near horizon geometry. 
The bosonic
symmetries arise from the isometries of $AdS_3\times S^3$.  The
isometries of the $AdS_3$ space form the non-compact group $SO(2,2)$,
while the isometries of $S^3$ form the group $SO(4)_E= SU(2)_E\times
\widetilde{SU(2)}_E$. The supergroups that contain this bosonic
subgroup $SO(2,2)\times SO(4)_E = (SL(2, R)\times SU(2)) \times {(SL(2,
  R) \times SU(2))}$ are either $Osp(3|2, R)\times Osp(3|2, R) $ and
$SU(1,1|2)\times SU(1,1|2)$. It is the latter that corresponds to the
symmetry of the $D_1-D_5$ system because the supercharges in this case
transform as a spinor of $ SO(4)_E$. We shall see that the
identification of the near horizon symmetry group $SU(1,1|2)\times SU(1,1|2)$
plays a crucial role in matching SCFT operators with the (dual) supergravity modes.

\subsection{Supergravity Solution with Non-zero vev of $B_{NS}$}

Our discussion so far has been devoted to SUGRA solutions in which the
values of all the moduli fields were set to zero. Such solutions have
the characteristic that the mass of the $D_1-D_5$ system is a sum of the 
charges that characterize the system. Such bound states are marginal, without any binding energy, and can
fragment into clusters of $D_1-D_5$ branes. The corresponding CFT has singularities.
In order to obtain a stable bound state and a non-singular CFT we have to turn on 
certain moduli fields. We will consider the case when $B_{NS}$ is non-zero.

The construction of  the supergravity solution that corresponds to a $\frac{1}{4}$ BPS
configuration, with a non-zero $B_{NS}$ was presented in \cite{DMWY}. 
See also \cite{russo}.
$B_{NS}$ has non-zero components only along the directions $6,7,8,9$ of the internal torus.
>From the view point of open string theory this is then a non-commutative torus.

Here we will
summarize the result. The solution contains, besides D1 and D5 brane charges, 
D3 brane charges that are induced by the $B_{NS}$. 
For simplicity we consider only non-zero values for $B_{79}$ and $B_{68}$. The 
asymptotic values are given by $B^{(\infty)}_{79} = b_{79}$ and $B^{(\infty)}_{68} = 
b_{68}$. It is important that at least 2 components of the $B_{NS}$ are non-zero, in order to be able to
discuss the self-dual and anti-self-dual components.

Below we present the full solution which can be derived by a solution generating 
technique. Details can be found in \cite{DMWY}.

\bea
\label{full-solution}
ds^2 &=& (f_1f_5)^{-1/2} (-dt^2 + (dx^5)^2) + (f_1f_5)^{1/2} (dr^2 +
r^2 d \Omega^2_3) \nonumber \\ [2mm]
&& + (f_1f_5)^{1/2} \left\{ Z^{-1}_\varphi ( (dx^6)^2 + (dx^8)^2) +
Z^{-1}_\psi ((dx^7)^2 + (dx^9)^2) \right\} , \\ [2mm]
\label{full-solution-1}
e^{2\phi} &=& f_1 f_5 / Z_\varphi Z_\psi , \\ [2mm]
\label{full-solution-2}
B^{(2)}_{NS} &=& (Z^{-1}_\varphi \sin\varphi \cos\varphi (f_1-f_5) +
b_{68})
dx^6 \wedge dx^8 \nonumber \\ [2mm]
&& + (Z^{-1}_\psi \sin\psi \cos\psi (f_1-f_5) + b_{79}) dx^7 \wedge
dx^9 , \\ [2mm]
\label{full-solution-3}
F^{(3)} &=& \cos\varphi \cos\psi \tilde K^{(3)} + \sin\varphi \sin\psi
K^{(3)} , \\ [2mm]
\label{full-solution-4}
F^{(5)} &=& Z^{-1}_\varphi (-f_5 \cos\varphi \sin\psi K^{(3)} + f_1
\cos\psi
\sin\varphi \tilde K^{(3)}) \wedge dx^6 \wedge dx^8 \nonumber \\ [2mm]
&&+ Z^{-1}_\psi (-f_5 \cos\psi \sin\varphi K^{(3)} + f_1 \cos\varphi
\sin\psi \tilde K^{(3)}) \wedge dx^7 \wedge dx^9 , \\ [2mm]
\label{full-solution-5}
Z_{\varphi, \psi} &=& 1 + {\mu_{\varphi,\psi} \over 2} \left({\alpha'
\over r^2}\right) , \ \ \ \mu_\varphi = \mu_1 \sin^2\varphi + \mu_5
\cos^2 \varphi , \ \ \ \mu_\psi =
\mu_1 \sin^2\psi + \mu_5
\cos^2 \psi.
\eea
Here $b_{68}$ and $b_{79}$ are arbitrary constants which we have added
at the end by a T-duality transformation that
shifts the NS B-field by a constant. Note that for $\varphi = \psi =
0$ and $b_{68} = b_{79} = 0$, the above solution reduces to the
known solution for $D_1-D_5$ system without B-field.

\vspace{2em}

The above solution depends upon 4 parameters $\mu_1$, $\mu_5$, and the
angles $\phi $ and $\psi $, and in general represents a system of D1,
D5 and D3 branes.  Since we are seeking a solution that has no source
D3 branes we require that the D3 brane charges are only induced by the
presence of the non-zero $B_{NS}$. This leads to certain conditions on
the solutions which we do not derive here, but whose physical
implication we analyze. We discuss both the asymptotically flat and 
near horizon geometry.

\vspace{5ex}
\noindent(i) {\it{Asymptotically Flat Geometry} }
\vspace{5ex}

In this case the induced D3 brane charges along the $(5,7,9)$ and  $(5,6,8)$ directions
are 
\be
\label{induced-d3}
Q_3 = B^{(\infty)}_{79} Q_5 , \ \ \ Q'_3 = B^{(\infty)}_{68} Q_5 ,
\ee
where
\be
\label{9}
B^{(\infty)}_{79} = b_{79} , \ \ \ B^{(\infty)}_{68} = b_{68} ,
\ee
There is a induced contribution to the D1 brane charge. The charge $Q_{1s}$ of 
the source D1 branes is 
\be
\label{induced-d1}
Q_{1s} = Q_1 - b_{68} \ b_{79} \ Q_5 .
\ee
while the D5 brane charge remains unaffected by the moduli.

\vspace{2ex}

{\noindent\it {Mass}}

\vspace{2ex}

Let us now study the mass formula as a function of the charges and the moduli.
The mass corresponding to the
${1\over 4}$ BPS solution
\cite{Obe-Pio98}, which coincides with the ADM mass, is given
in terms of the appropriate charges by
\be
\label{10}
M^2 = (Q_1 + Q_5)^2 + (Q_3 - Q'_3)^2
\ee
This can in turn be expressed in terms of $Q_{1s}$, $Q_5$ and $b_{68}, b_{79}$
\be
\label{mass-flat-mod}
M^2 = (Q_{1s}+b_{68}b_{79}Q_5 + Q_5)^2 + Q_5^2 (b_{68} - b_{79})^2
\ee
We must consider the mass as a function of the moduli, holding $Q_{1s}$ and  $Q_5$
fixed. 
We see that for non-zero moduli we have a true bound state that turns marginal when the 
moduli are set to zero. To locate the values of the moduli which minimize the mass, we extremize the mass w.r.t the moduli. The extremal values of the moduli are
\be
\label{12}
b_{68} = - b_{79} = \pm \sqrt{Q_{1s}/Q_5 -1}  ,
\ee
This says that the $B_{NS}$ moduli are self-dual, in the asymptotically flat metric.
The mass at the critical point of the true bound state is then given by
\be
\label{13}
M^2 = 4 Q_{1s} Q_5 
\ee
\vspace{2ex}

{\noindent\it {Near Horizon Geometry}}

\vspace{2ex}

In this case, absence of D3-brane sources is ensured if we set
\be
\label{no-d3-nh}
Q^{(h)}_3 = B^{(h)}_{79} Q_5 , \ \ \ Q^{(h)'}_3 = B^{(h)}_{68} Q_5 ,
\ee
where
\bea
\label{14a}
B^{(h)}_{68} &=& {\mu_1 - \mu_5 \over \mu_\varphi} \sin\varphi
\cos\varphi + b_{68} , \\ [2mm]
\label{14b}
B^{(h)}_{79} &=& {\mu_1 - \mu_5 \over \mu_\psi} \sin\psi
\cos\psi + b_{79} ,
\eea
are the horizon values of the two nonzero components of the
B-field. 
Moreover, we see can that
in this case
\be
\label{15}
{B^{(h)}_{68}\over \mu_\psi} = - {B^{(h)}_{79} \over \mu_\varphi} ,
\ee
which is the self-duality condition on the B-field in the near horizon
geometry. We also note that the volume of $T^4$ at the horizon is given
by
\be
\label{16}
V^{(h)}_{T^4} = {\mu_1\mu_5 \over \mu_\varphi \mu_\psi} = {Q^{(h)}_1s
\over Q_5} .
\ee

The D1-brane charge that arises from
source D1-branes in this case is given by
\be
\label{17}
Q^{(h)}_{1s} = Q^{(h)}_1 - B^{(h)}_{68} B^{(h)}_{79} Q_5 .
\ee
One can show that
\be
\label{18}
Q^{(h)}_{1s} = Q_{1s}
\ee
where $Q_{1s}$ is given by (\ref{17}). Thus we see that not only do the
parameters $b_{68}$ and $b_{79}$ have the same values here as in the
asymptotically flat case, even the source D1-branes are identical,
despite the total D1-brane charges being very different in the two
cases.

\vspace{2ex}

\noindent{\it {Mass}}

\vspace{2ex}
The ${1\over 4}$ BPS mass formula in terms of the various charge
densities in this case is
\be
\label{19}
\left({M^{(h)} \over V^{(h)}_{T^4}}\right)^2 = \left({Q^{(h)}_1 \over
V^{(h)}_{T^4}} + Q_5\right)^2 + \left({Q^{(h)}_3\over
\sqrt{g_{77}g_{99}}} - {Q^{(h)'}_3 \over \sqrt{g_{66} g_{88}}}\right)^2
\ee
Using (\ref{no-d3-nh})-(\ref{18})
it can be easily seen that
\be
\label{20}
\left(M^{(h)}\right)^2 =  V^{(h)}_{T^4} \left(4Q_{1s}Q_5\right).
\ee
Apart from the extra factor of the $T^4$ volume in the near horizon
geometry, this is exactly the same as \eq{13}. The extra volume factor
correctly takes into account the difference in the 6-dimensional
Newton's constant between the asymptotically flat and near horizon
geometries because of the difference in the $T^4$ volume in the two
cases. We have already seen that the B-field is automatically
self-dual in the near horizon geometry and that the volume of $T^4$
satisfies the condition given by \eq{16} and \eq{17}. We now see that
the mass of the bound state is already at the fixed point value. Thus
the solution we
have here provides an explicit demonstration of the attractor mechanism
\cite{Ferrara}.

The significance of this solution is that it is the description of a stable bound state
in the near horizon geometry. As we shall discuss later this situation corresponds to
a non-singular dual CFT. 

\subsection{Semi-Classical Absorption Cross-Section and Semi-Classical \\
Hawking Radiation Formula}
Now that we have discussed the various classical solutions we want to summarize
the basic steps in the calculation of the semi-classical 
absorption cross-section and it's relation to the emission rate of Hawking 
radiation from a black hole \cite {Dha-Man-Wad96,Das-Mat96,Mal-Str96}. We do the calculation for minimal scalars in the s-wave.
These fields satisfy a linear equation in which only the Einstein metric is present, 
leading to a great simplification in the calculation. 
\be
D_\mu \del^\mu \varphi =0
\ee
For the 5-dim. black hole discussed earlier the s-wave radial equation becomes

\be
[\frac{h}{r^3}\frac{d}{dr} (h r^3 \frac{d}{dr}) + f w^2]R_w(r) =0
\ee
where $f=f_1f_5$ and 
\be
\varphi= R_w(r) \exp[-iwt]
\ee
Introducing  $\psi = r^{3/2} R$ 
and $r_* = r + \frac{r_o}{2}\ln |\frac{r-r_o}{r+r_o}|$
we have the Schroedinger type equation
\be
[-\frac{d^2}{dr^2_*}  + V_w(r_*)]\psi = 0
\ee
where
\be
V_w (r_*) = -w^2 f + 
\frac{3}{4 r^2}(1 + 2 r_0^2/r^2 - 3 r_0^4/r^4)
\ee
The basic idea is to solve the equation in 2 regions with appropriate
boundary conditions and then match the solution in the overlapping
region.
In order to do so we need to choose the parameters characterizing the solution to be in the following range,
\bea
r_0, r_n &\ll&  r_1,r_5
\nn\\
wr_5 &\ll& 1
\nn\\
r_1\sim r_5,\, r_0\sim r_n 
\eea
Also the wave length of the incident radiation $1/w$ is comparable to the thermal wavelength specified by the $1/T_H$.
The far and near solutions are matched at a point 
$r_m$ such that
\be
r_0, r_n \ll r_m \ll r_1, r_5, \qquad wr_1 \ll r_m/r_1
\ee

\vspace{3ex}

{\it {Far Zone ($r \ge r_m$):} }

\noindent Here the potential $V_w$ becomes (in terms of 
$\rho=wr$)
\be
V_w(\rho)= -w^2(1 - \frac{3}{4 \rho^2})
\ee
This is Bessel's equation, so that
\bea
\psi &=& \alpha F(\rho) + \beta G(\rho)
\nn\\
F(\rho) &=& \sqrt{\pi \rho/2}J_1(\rho),\quad 
G(\rho) = \sqrt{\pi \rho/2}N_1(\rho)
\eea
For $\rho\to \infty$ one can easily see the coefficients of the 
incoming wave $e^{-iw r}$ and the outgoing wave
$e^{iw r}$.

\vspace{3ex}
{\it{Near zone ($r \le r_m$):}}

\ni This is the region near the pit of the potential, or the throat region. 
Here we get a hyper-geometric equation,
\be
\frac{h}{r^3}\frac{d}{dr} (h r^3 \frac{d}{dr}R) 
+ [\frac{(wr_nr_1r_5)^2}{r^6}
+ \frac{w^2r_1^2 r_5^2}{r^4} ]R_w(r) =0
\ee
which is solved by 
\bea
R &=& AR_{in}+ BR_{out} 
\nn\\
R_{in} &=& z^{-i(a+b)/2} F(-ia, -ib, 1-ia -ib,z)
\nn\\
R_{out }&=& z^{i(a+b)/2} F(-ia, -ib, 1-ia -ib,z)
\nn\\
z &=& (1- r_0^2/r^2)
\nn\\ 
a &=& w/(4 \pi T_R), b=w/(4 \pi T_L)
\nn\\
\eea
The temperatures $T_{R,L}$ are given by
\be
T_{L,R} = \frac{r_0}{2\pi r_1 r_5}e^{\pm \sigma}
\ee
The important boundary condition that we impose is $B=0$. This says that at the black hole horizon there is 
no outgoing wave. 

\vspace{3ex}

$R$ and $\frac{d}{dr} R$ can now be matched in the overlapping region (below the potential barrier) at some point $r_m$. The matching conditions imply
\bea
\sqrt{\pi/2} w^{3/2} \alpha/2 &=& A e_1
\nn\\
e_1 &\equiv& \frac{\Gamma(1-ia-ib)}{\Gamma(1-ib)\Gamma(1-ia)}
\nn\\
\beta/\alpha &\ll & 1
\eea

\vspace{3ex}

Now that we have constructed the solution we can calculate the flux from the Schroedinger equation
\be
{\cal F}(r) = \frac{1}{2i}[ R^* hr^3 dR/dr - {\rm c.c.}]
\ee
This flux is `conserved' $\frac{d}{dr} {\cal F}=0$.

The fraction of the flux that gets absorbed at the horizon is given by the ratio of the flux calculated from the solution at the horizon (where we used the horizon boundary condition) and the flux due to the incoming spherical wave from infinity,
\be
R_1 = {\cal F}(r_0)/{\cal F}^{in}(\infty)= r_0^2 
\frac{a+b}{w|e_1|^2} w^3 \pi/2
\ee

\newpage

\ni{\bf Absorption Cross-section:}

\vspace{5ex}
To calculate the absorption cross-section of an incident plane wave as opposed to the spherical wave that we did the above calculation with, we have to introduce a conversion factor. This is easily done by the expansion of a plane wave in terms of spherical waves,
\be
e^{-iwz} = (4\pi/w^3) e^{-iwr}Z_{000} + {\rm other\;partial\;waves}
\ee
Taking this into account we get \cite{Mal-Str96}
\bea
\label{class-abs}
\sigma_{abs} &=& (4\pi/w^3) R_1
\nn\\
\; &=& 2 \pi^2 r_1^2 r_5^2 \frac{\pi w}{2} 
\frac{\exp(w/T_H)-1}{(\exp(w/2T_R)-1) (\exp(w/2T_L)-1)}
\eea
In the $w\to 0$ limit, one gets \cite{Dha-Man-Wad96}
\be
\label{abs}
\sigma_{abs} = A_h
\ee
where $A_h$ denotes the area of the event horizon.

The decay rate is given by the well known formula of Hawking,
\be
\Gamma = {\rm Prob}_{\rm decay}\frac{V_4}{{\tilde R T}} 
\frac{d^4k}{(2 \pi)^4}, \,\, {\tilde R}=Q_1Q_1R
\ee
giving
\be
\label{decay}
\Gamma_H = \sigma_{abs} (e^{w/T_H}-1)^{-1} 
\frac{d^4k}{(2 \pi)^4}
\ee
\vspace{3ex}

With this we conclude our discussion of supergravity aspects and now turn to explaining some of the important thermodynamical
formulas from the viewpoint of string theory.

\section{\bf Microscopic Modeling of the Black Hole in terms of the $D_1-D_5$ System}

Our aim here is to study the low energy collective excitations of the $D_1-D_5$ system.
There are 2 ways to proceed and we shall discuss both of them. The first method is 
a description in terms of a 2-dim. gauge theory and the second
method involves identifying D1 branes with instantons of a 4 dim. gauge theory.
The latter description is more accurate and is valid for instantons of all sizes.
The 2-dim. gauge theory description is valid for small instanton size but it is more
physical and gives a feeling for the dynamics. We will discuss this more approximate
description first.

\section{\bf The $D_1/D_5$ System and the ${\cal N}=4$,\,$U(Q_1)\times U(Q_5)$ gauge theory in 2-dimensions }

Consider type IIB string theory with five coordinates, say 
$x^5\cdots x^9$,
compactified on $S^1\times T^4$. The microscopic model consists $Q_1$ D1-branes and
$Q_5$ D5-branes \cite {Str-Vaf96,Cal-Mal96}. The D1-branes are parallel
to the $x^5$ coordinate compactified to a circle $S^1$ of radius $R$,
while the D5-branes are parallel to $x^5$ and $x^6,\cdots,x^9$
compactified on a torus $T^4$ of volume $V_4$. The charge $N$ is
related to the momenta of the excitations of this system
along $S^1$. We take the $T^4$ radii to be of the order of $\alpha'$
and smaller than $R$ which, in turn, is much smaller than the black
hole radius. 

We shall see that the low-energy dynamics of this D-brane system is
described by a $U(Q_1)\times U(Q_5)$ gauge theory in two dimensions
with $N=4$ supersymmetry \cite {Mal96,Has-Wad97b}. The gauge theory
will be assumed to be in the Higgs phase because we are interested in
the bound state where the branes are not seperated from each other in
the transverse direction. In order to really achieve this and prevent
branes from splitting off we will turn on the Fayet-Illiopoulos
parameters. In supergravity these correspond to the vev of the
Neveu-Schwarz $B_{NS}$. In principle we can also turn on the $\theta $
term in the gauge theory. This corresponds to a vev of a certain linear combination of the RR $0$-form and $4$-form.

The elementary excitations of the D-brane system correspond to open
strings with two ends attached to the branes and there are three
classes of such strings: the (1,1), (5,5) and (1,5) strings. The
associated fields fall into vector multiplets and hypermultiplets,
using the terminology of N=2, D=4 supersymmetry. 
\gap

\noindent
{\bf $(1,1)$ strings} 

The part of the spectrum coming from (1,1) strings is
simply the dimensional reduction, to $1+1$ dimensions (the
$(t,x^5)$-space), of the N=1, $U(Q_1)$ gauge theory in $9+1$
dimensions \cite{Pol-tasi96}.  

The bosonic fields of this theory can be 
organized
into the vector multiplet and the hypermultiplet of $N=2$ theory in
four-dimensions as
\bea
\mbox{Vector multiplet:} \; A_0^{(1)}, A_5^{(1)}, Y_m^{(1)}, m=1,2,3,4
\\   \nonumber
\mbox{Hypermultiplet:} \; Y_i^{(1)}, i=6,7,8,9
\eea
The $A_0^{(1)}, A_5^{(1)}$ are the $U(Q_1)$ gauge fields in the
non-compact directions. The $Y_m^{(1)}$'s and $Y_i^{(1)}$'s 
are gauge fields in the compact directions of the $N=1$
super Yang-Mills in ten-dimensions. 
They are hermitian $Q_1\times Q_1$ matrices transforming as
adjoints of
$U(Q_1)$. The hypermultiplet of $N=2$ supersymmetry are
doublets of the $SU(2)_R$ symmetry of the theory. 
The adjoint matrices $Y_i^{(1)}$'s can be arranged as 
doublets under $SU(2)_R$ as
\be
N^{(1)} = 
\left(
\begin{array}{c}
N_1^{(1)} \\ \nonumber
N_2^{(1)\dagger}
\end{array}
\right)
=\left(
\begin{array}{c}
Y_9^{(1)} + i 
Y_8^{(1)}  \\
Y_7^{(1)} - i 
Y_6^{(1)} 
\end{array}
\right)
\ee

\gap

\noindent
{\bf $(5,5)$ strings} 

The field content of these massless open 
strings is similar to the the $(1,1)$ strings except for 
the fact that the gauge group is 
$U(Q_5)$ instead of $U(Q_1)$.
Normally one would have expected
the gauge theory of the $(5,5)$ strings to be a 
dimensional reduction of $N=1$ $U(Q_5)$ super Yang-Mills to
$5+1$ dimensions. Since we are ignoring the Kaluza-Klein modes 
on $T^4$ this is effectively a theory in $1+1$ dimensions. The vector
multiplets and the hypermultiplets are given by
\bea
\mbox{Vector multiplet:} \; A_0^{(5)}, A_5^{(5)}, Y_m^{(5)} \, m=1,2,3,4
\\   \nonumber
\mbox{Hypermultiplet:} \; Y_i^{(5)} \, i=6,7,8,9
\eea
The $A_0^{(5)}, A_5^{(5)}$ are the $U(Q_5)$ gauge fields in the
non-compact directions. The $Y_m^{(5)}$'s and $Y_i^{(5)}$'s 
are gauge fields in the compact directions of the $N=1$
super Yang-Mills in ten-dimensions. They 
are hermitian $Q_5\times Q_5$ matrices transforming as adjoints of
$U(Q_5)$. The hypermultiplets $Y_i^{(5)}$'s can be 
arranged as doublets under
$SU(2)_R$ as
\be
N^{(5)} = 
\left(
\begin{array}{c}
N_1^{(5)} \\ \nonumber
N_2^{(5)\dagger}
\end{array}
\right)
=
\left(
\begin{array}{c}
Y_9^{(5)} + i 
Y_8^{(5)}  \\
Y_7^{(5)} - i 
Y_6^{(5)} 
\end{array}
\right)
\ee

Since $x^m$ are compact, the (1,1) strings can also have winding modes
around the $T^4$. These are, however, massive states in the
$(1+1)$-dimensional theory and can be ignored. This is because their
masses are proportional to $R \gg \sqrt{\alpha '}$. Similarly, the
part of the spectrum coming from (5,5) strings is the dimensional
reduction, to $5+1$ dimensions, of the N=1, $U(Q_5)$ gauge theory in
$9+1$ dimensions. In this case, the gauge field components $A^{(5)}_m$
($m=6,7,8,9$) also have a dependence on $x^m$. Momentum modes
corresponding to this dependence are neglected because the size of the 4-torus is of the order of the string scale $\sqrt{\alpha '}$. 
The neglect of the winding modes of the $(1,1)$ strings and the KK modes of the $(5,5)$ strings is consistent with T-duality.
A set of four T-duality
transformations along $x^m$ interchanges D1- and D5-branes and also
converts the momentum modes of the (5,5) strings along $T^4$ into
winding modes of (1,1) strings around the dual torus \cite{WT}. Since
these winding modes have been ignored, a T-duality covariant
formulation requires that we should also ignore the associated
momentum modes.

\gap

\noindent
{\bf $(1,5)$ and $(5,1)$ strings} 

The field content obtained so far is that of N=2, \, $U(Q_1)\times U(Q_5)$
gauge theory, in 1+5 dimensions, reduced to 1+1 dimensions on $T^4$.

The $SO(4)\sim SU(2)_L\times SU(2)_R$ rotations on the tangent space
of the torus act on the components of the adjoint hypermultiplets
$X^{(1,5)}_m$ as an $R$-symmetry. To this set of fields we have to add
the fields from the (1,5) sector that are constrained to live in 1+1
dimensions by the ND boundary conditions. These strings have their
ends fixed on different types of D-branes and, therefore, the
corresponding fields transform in the fundamental representation of
both $U(Q_1)$ and $U(Q_5)$.  The ND boundary conditions have the
important consequence that the (1,5) sector fields form a
hypermultiplet which is chiral w.r.t. $SO(4)_I$.  The chirality
projection is due to the GSO projection. Hence the $R$-symmetry group
is $SU(2)_R$.
\be
\chi_{a\bar{b}} = 
\left(
\begin{array}{c}
A_{a\bar{b}} \\
B^{\dagger}_{a\bar{b}}
\end{array}
\right)
\ee
A few comments are in order:
\begin{enumerate}
\item The inclusion of these
fields breaks the supersymmetry by half, to the equivalent of N=1 in
D=6, and the final theory only has $SU(2)_R$ $R$-symmetry. 

\item The fermionic superpartners of these
hypermultiplets which arise from the Ramond sector of the massless
excitations of
$(1,5)$ and $(5,1)$ strings carry spinorial indices under $SO(4)_E$
and they are singlets under $SO(4)_I$. 

\item The $U(1) \times
\overline{U(1)}$ subgroup is important. One combination leaves the
hypermultiplet invariant. The other combination is active and
$(A_{a'a}, B_{a'a})$ have $U(1)$ charges $(+1, -1)$. 

\item $\chi$ is a
chiral spinor of $SO(4)_I$ with convention $\Gamma_{6789} \ \chi = -
\chi$.

\item Since we are describing the Higgs phase in which all the branes sit on top
of each other we have $Y_i^{(1,5)}=0$. 

\item In the above discussion, the fields $Y_{i}$ and $X_{i}$ along the
torus directions are assumed to be compact. However it is not obvious
how to compactify the range of $\chi$ so that the integration over
this field in the path integral is finite.
\end{enumerate}

In summary, the gauge theory of the $D_1-D_5$ system is a $1+1$ dimensional 
$(4,4)$ supersymmetric gauge theory with gauge group $U(Q_1)\times
U(Q_5)$. The matter content of this theory consists of hypermultiplets
$Y^{(1)}$'s, $Y^{(5)}$'s transforming as adjoints of $U(Q_1)$ and
$U(Q_5)$ respectively. It also has the hypermultiplets $\chi$'s which
transform as bi-fundamentals of $U(Q_1)\times \overline{U(Q_5)}$. 

\subsection{The Potential Terms}

The lagrangian of the above gauge theory can be worked out from the dimensional reduction of $d=4$,\, ${\cal N}=2$ gauge theory.The potential energy density of the vector and hyper multiplets is a
sum of 4 positive terms. In this section for convenience of notation 
we have defined
$Y_i^{(1)}=Y_i$, $Y_i^{(5)}=X_i$, $Y_m^{(1)}=Y_m$, $Y_m^{(5)}=X_m$)
\bea
\label{v}
V &=& V_1 + V_2 + V_3 + V_4 \\ [2mm]
\label{v.1}
V_1 &=& - {1\over 4g^2_1} \sum_{m,n} tr_{U(Q_1)} [Y_m, Y_n]^2 - {1\over
4g^2_5} \sum_{m,n} tr_{U(Q_5)} [X_m, X_n]^2 \\ [2mm]
\label{v.2}
V_2 &=& - {1\over 2g^2_1} \sum_{i,m} tr_{U(Q_1)} [Y_i, Y_m]^2 - {1\over
2g^2_5} \sum_{i,m} [X_i, X_m]^2 \\ [2mm]
\label{v.3}
V_3 &=& {1\over 4} \sum_m tr_{U(Q_1)} (\chi X_m - Y_m \chi) (X_m
\chi^\dagger -
\chi^\dagger Y_m)^2 \\ [2mm]
V_4 &=& {1\over 4} tr_{U(Q_1)} (\chi i \Gamma^T_{ij} \chi^+ + i [Y_i,
Y_j]^+ - \zeta^+_{ij} {1\!\!\!1 \over Q_1})^2 \nonumber \\ [2mm]
\label{v.4}
&& + {1\over 4} tr_{U(Q_5)} (\chi^+ i \Gamma_{ij} \chi + i [X_i,
X_j]^+ - \zeta^+_{ij} {1\!\!\!1 \over Q_5})^2
\eea
The potential energy $V_4$ comes from a combination of $F$ and $D$
terms of the higher dim. gauge theory. $\Gamma_{ij} = {i\over 2}
[\Gamma_i, \Gamma_j]$ are spinor rotation matrices. The notation
$a^+_{ij}$ denotes the self-dual part of the anti-symmetric tensor
$a_{ij}$.

In $V_4$ we have included the Fayet-Iliopoulos (FI) terms
$\zeta^+_{ij}$, which form a triplet under $SU(2)_R$. Their inclusion
is consistent with $N=4$ SUSY. The FI terms can be identified with the
self dual part of $B_{ij}$, the anti-symmetry tensor of the NS sector
of the closed string theory \cite{Sei-Wit99}. This identification at this stage rests
on the fact that (i) $\zeta^+_{ij}$ and $B^+_{ij}$ have identical
transformation properties under $SU(4)_I$ and (ii) at the origin of
the Higgs branch where $\chi = X = Y = 0$, $V_4 \sim \zeta^+_{ij}
\zeta^+_{ij}$. This signals a tachyonic mode from the view point of
string perturbation theory. The tachyon mass is easily computed and
this implies the relation $\zeta^+_{ij} \zeta^+_{ij} \sim B^+_{ij}
B^+_{ij}$.

\subsection{D-Flatness Equations and the Moduli Space}

The supersymmetric ground state (semi-classical) is characterized by
the 2-sets of D-flatness equations which are obtained by setting $V_4
= 0$. They are best written in terms of the $SU(2)_R$ doublet fields
$N^{(1)}_{a'b'}$ and $N^{(5)}_{ab}$ :
\bea
N^{(1)} &=& \pmatrix{N^{(1)}_1 \cr N^{(1)}_2} = \pmatrix{Y_9 + i Y_8
\cr Y_7 + i Y_6} \nonumber \\ [2mm]
\label{3.7}
N^{(5)} &=& \pmatrix{N^{(5)}_1 \cr N^{(5)}_2} = \pmatrix{X_9 + iX_8
\cr X_7 + iX_9}
\eea

We also define $\zeta=\zeta^{+}_{69}$ and $\zeta_{c}=
\zeta^{+}_{67}+ i\zeta^{+}_{68}$.
With these definitions the 2 sets of D-flatness conditions become:
\be
\label{D-terms}
(AA^+ - B^+B)_{a'b'} + [N^{(1)}_1, N^{(1)\dagger}_1]_{a'b'} -
[N^{(1)}_2,
N^{(1)\dagger}_2]_{a'b'} = {\zeta \over Q_1} \delta_{a'b'}
\ee
\be
(AB)_{a'b'} + [N^{(1)}_1, N^{(1)\dagger}_2]_{a'b'} = {\zeta_{c} \over
Q_1} \delta_{a'b'}
\ee
\be
(A^+A - BB^+)_{ab} + [N^{(5)}_1, N^{(5)\dagger}_1]_{ab} - [N^{(5)}_2,
N^{(5)\dagger}_2]_{ab} = {\zeta \over Q_5} \delta_{ab}
\ee
\be
(A^+B^+)_{ab} + [N^{(5)}_1, N^{(5)\dagger}_2]_{ab} = {\zeta_{c} \over
Q_5} \delta_{ab}
\ee

The hypermultiplet moduli space is a solution of the above equations
modulo the gauge group $U(Q_1) \times U(Q_5)$. A detailed discussion
of the procedure was given in \cite {Has-Wad97b,DMWY}. Here we summarize.

If we take the trace parts of (\ref{D-terms})
we get the {\em same} set of 3
equations as the D-flatness equations for  a $U(1)$ theory
with $Q_1 Q_5$ hypermultiplets, with $U(1)$ charge assignment $(+1,
-1)$ for $(A_{a'b}, B^T_{a'b})$. Thus,
\bea
\label{Tcpn1}
\sum_{a'b} (A_{a'b} A^\ast_{a'b} - B^T_{a'b} B^{T\ast}_{a'b}) = \zeta
\eea
\be
\label{Tcpn2}
\sum_{a'b} A_{a'b} B^T_{a'b} = \zeta_c
\ee
For a given point on the surface defined by 
(\ref {Tcpn1}),(\ref{Tcpn2})
the traceless parts of (\ref{D-terms}) lead to $3Q^2_1 + 3Q^2_5 - 6$
constraints 
among $4Q^2_1 + 4Q^2_5 - 8$ degrees of freedom corresponding to the
traceless parts of the adjoint hypermultiplets $N^{(1)}$ and
$N^{(5)}$. Using $Q^2_1 + Q^2_5 - 2$ gauge conditions corresponding to
$SU(Q_1) \times SU(Q_5)$ we have $(3Q^2_1 + 3Q^2_5 - 6) + (Q^2_1 +
Q^2_5 - 2) = 4Q^2_1 + 4Q^2_5 - 8$ conditions for the $(4Q^2_1 + 4Q^2_5
-8)$ degrees of freedom in the traceless parts of $N^{(1)}$ and
$N^{(5)}$. The 8 degrees of freedom corresponding to $trX_i$ and
$trY_i$, $i = 6,7,8,9$ correspond to the centre-of-mass
of the D5 and D1 branes
respectively.

\vspace{1em}

\subsection{The Bound State in the Higgs Phase}

Having discussed the moduli space that characterizes the SUSY ground
state we can discuss the fluctuations of the transverse vector
multiplet scalars $X_m$ and $Y_m$, $m = 1,2,3,4$. In the Higgs phase
since $\langle X_m \rangle = \langle Y_m \rangle = 0$ and $\chi =
\overline \chi$ lies on the surface defined by (\ref{Tcpn1}),(\ref{Tcpn2}).
The relevant
action of fluctuations in the path integral is,
\be
S = \sum_m \int dt dx_5 (tr_{U(Q_5)} \partial_\alpha X_m
\partial^\alpha X_m + tr_{U(Q_1)} \partial_\alpha Y_m \partial^\alpha
Y_m) + \int dt dx_5 (V_2 + V_3)
\ee

We restrict the discussion to the case when $Q_5=1$ and $Q_1$ is
arbitrary.
In this case the matrix $X_m$ is a real number which we denote by $x_m$.

$\chi$ is a complex column vector with components $(A_{a'},B_{a'})$,
$a'= 1,..
.,Q_1$. Since we are looking at the fluctuations of the $Y_m$ only to
quadratic
order in the path integral, the integrals over the different $Y_m$
decouple
from each other and we can treat each of them separately. Let us discuss
the
fluctuation $Y_1$ and set $(Y_1)_{a'b'}=\delta_{a'b'} y_{1{a'}}$. Then
the potential
$V_3$, (\ref{v.3}) becomes
\bea
V_3= \sum_{a'} (|A_{a'}|^2 + |B_{a'}|^2 )(y_{1a'} -x_1)^2
\eea
We will prove that $|A_{a'}|^2 + |B_{a'}|^2$ can never vanish if the FI
terms
are non-zero.
In order to do this let us analyze the complex D-term equation (\ref{Tcpn2})
\bea
\label{Tcpn-3}
A_{a'}B_{b'}+ [N^{(1)}_1,N^{(1)\dagger}_2]_{a'b'} = {\zeta_c \over Q_1}
\delta_{a'b'}
\eea
We can use the complex gauge group $GL(C,Q_1)$ to diagonalize the
complex
matrix $N^{(1)}_1$ \cite {witt-higgs-br}. Then, (\ref{Tcpn-3}) becomes
\bea
A_{a'}B_{b'}+ (n_{a'} - n_{b'})(N^{(1)\dagger}_2)_{a'b'} ={\zeta_c \over
Q_1}
\delta_{a'b'}
\eea
For $a' \neq b'$, this determines the non-diagonal components of
$N^{(1)}_2$
\bea
(N^{(1)\dagger}_2)_{a'b'} = - { A_{a'}B_{b'} \over n_{a'} - n_{b'} }
\eea
For $a=b$, we get the equations
\bea
A_{a'}B_{a'}={\zeta_c \over Q_1} , a'= 1,..,Q_1
\eea
which imply that
\bea
|A_{a'}| |B_{a'}| = { |\zeta_c| \over Q_1}
\eea
with the consequence that $|A_{a'}|$ and  $|B_{a'}|$ are non-zero for
all
$a' = 1,..,Q_1$. This implies that $(|A_{a'}|^2 + |B_{a'}|^2) > 0)$, and
hence
the fluctuation $(y_{1a'} - x_1)$ is massive. If we change variables
$y_{1a'}
\rightarrow y_{1a'} +x_1$, then $x_1$ is the only flat direction. This
corresponds to the global translation of the 5-brane in the $x_1$
direction.

A similar analysis can be done for all the remaining directions
$m=2,3,4$ with
identical conclusions. This shows that a non-zero FI term implies a true
bound
state of the $Q_5=1$, $Q_1=N$ system.
If $FI=0$, then there is no such guarantee and the
system can easily fragment, due to the presence of flat directions in
$(Y_m)_{a'b'}$. 

What the above result says is that when the FI parameters are non-zero
the zero mode of the fields $(Y_m)_{a'b'}$
is massive. If we regard the zero mode as a collective coordinate then
the Hamiltonian of the zero mode has a quadratic potential which agrees
with
the near horizon limit of the Liouville potential derived in 
\cite {Sei-Wit99,DMWY}.

The general case with an arbitrary number of $Q_1$ and $Q_5$ branes
seems significantly harder to prove, but the result is very plausible on physical grounds. If the potential for a single test $D1$ brane is attractive, it is hard to imagine any change in this fact if there are 2 test $D1$
branes, because the $D1$ branes by themselves can form a bound state.

\subsection{The Conformally Invariant Limit of the Gauge Theory}

The sigma model that describes the low energy modes corresponding to
the hyper-multiplet moduli defined by the equations (\ref{D-terms}) is
given by the lagrangian (bosonic part), 
\be
\label{sigmamodel}
S = \sum_m \int dt dx_5 (tr_{U(Q_5)} \partial_\alpha X_i
\partial^\alpha X_i + tr_{U(Q_1)} \partial_\alpha Y_i \partial^\alpha
Y_i) + \int dt dx_5 (\partial_\alpha \chi \partial^\alpha \chi^{\dagger})
\ee
This is a very difficult non-linear system, with $N=4$ SUSY. Since we
are interested in the low energy dynamics we may ask whether there is
a SCFT fixed point.  Such a SCFT must have (4,4) supersymmetry (16 real supersymmetries) with a
central charge $c=6(Q_1Q_5 +1)$.  Now note that the equations
(\ref{Tcpn1}),(\ref{Tcpn2}) describe a hyper-Kahler manifold and hence the
sigma model defined on it is a SCFT with (4,4) SUSY. We can then
consider the part of the action involving the $X_i$ and $Y_i$ which
are solved in terms of the $\chi $ as giving a deformation of the
SCFT. Now this deformation clearly reduces the SUSY to $N=4$, but
seems to preserve the original degrees of freedom. For this reason the
deformation can be identified with a set of marginal or irrelevant
operators. Inspite of the simplification at the fixed point this
theory is difficult to work with. 

The sigma model action at the conformally invariant point is 
\be
\int dt dx_5\sum_{a'b} (\del_\alpha A_{a'b} \del_\alpha A^\ast_{a'b} -\del_\alpha B^T_{a'b}\del_\alpha B^{T\ast}_{a'b})
\ee
The sigma model fields are constrained to be on the surface defined by 
(\ref{Tcpn1}),(\ref{Tcpn2}). Further after appropriate gauge fixing the residual gauge invariance inherited from the gauge theory is the Weyl group $S(Q_1)\times S(Q_5)$ \cite{Has-Wad97b}.
The Weyl invariance can be used to construct gauge invariant strings of various lengths. If $Q_1$ and $Q_5$  are relatively prime it is indeed possible to prove the existence of a single winding string with minimum unit of momentum given by $\frac{1}{Q_1Q_5}$. This is associated with the longest cyclic subgroup of $S(Q_1)\times S(Q_5)$. Cyclic subgroups of shorter length cycles lead to strings with minimum momentum $\frac{1}{l_1l_5}$, where
$l_1$ and $l_5$ are the lengths of the cycles. In a different way of describing these degrees of freedom we shall see in the next sections that strings of various lengths are associated with chiral primary operators of the conformal field theory on the moduli space of instantons on a 4-torus.

We conclude this section by showing that certain deductions about thermodynamic properties can be made just by the
knowledge of the central charge and the level of the Virasoro
algebra. This information is sufficient to calculate the number of micro-states.  To find the
microstates of the $D_1-D_5$ black hole we look for
states with $L_0 =N_L$ and $\bar{L}_0 = N_R$.  The assymptotic number
of distinct states of this SCFT is given by Cardy's formula
\be
\Omega = \exp (2\pi( \sqrt{Q_1Q_5 N_L}+ \sqrt{Q_1Q_5 N_R}))
\ee
>From the Boltzmann formula one obtains
\be
\label{microentropy}
S=2\pi( \sqrt{Q_1Q_5 N_L}+\sqrt{Q_1Q_5 N_R})
\ee
This exactly reproduces the Bekenstein-Hawking entropy. For the
extremal (BPS) case, $N_R=0$, and for the near extremal case $N_L=N+n$
and $N_R=n$, where $n<<N$. For the near extermal cases 
(\ref{microentropy}) also gives the correct Hawking temperature $T_H$
\bea
&& T_L^{-1}  =  \frac {1}{R}\frac {\del S}{\del N_L}  =  \frac {\pi}{R}\sqrt \frac{Q_1Q_5}{N_L} \nn \\
&& T_R^{-1}  =  \frac {1}{R}\frac {\del S}{\del N_R}  =  \frac {\pi}{R}\sqrt \frac{Q_1Q_5}{N_R}  \nn \\
&& T_H^{-1}  =  \frac {1}{2}(T_L^{-1}+T_R^{-1})
\eea

\section{\bf $D_1$ Branes as Instantons of the $D_5$ Gauge Theory}
In this section we take a different approach to the
description of $D_1$ branes \cite{Dou95}. We will see that we can find D1 branes within D5 branes!
We begin with a theory of $Q_5$, $D_5$
branes along the compact coordinates $x_i, i=5,6,7,8,9$. The low
energy degrees of freedom of this system are described by a N=2, $U(Q_5)$ 
gauge theory in 6 dimensions. This gauge theory has a
dimensional coupling constant $g_6$, and hence it is not
renormalisable. This means that it cannot capture degrees of freedom
at the string scale and hence is valid for wavelenghts much larger than
the string scale which acts as a short distance cutoff.

In this gauge theory let us look for configurations which break
the 16 supersymmetries to 8. The reason is that we know that the
presence of $D_1$ branes would do exactly that. Further since the
$D_1$ branes are strings moving in time along the $x_5$ direction and
smeared all over the 4-torus ($x_i, i=6,7,8,9$) we look for gauge field
configurations which, to begin with, depend only on the torus coordinates.
Such configurations are well known and easy to find.

We consider the variation of the gaugino under a supersymmetry transformation
and set it to zero
\be
\delta_\epsilon \lambda = \Gamma_{ab} F^{ab} \epsilon =0
\ee
where $a,b$ run over $6,7,8,9$. It is easy to see that
this is equivalent to
\be
\label{instanton-equation}
F_{ab} = \epsilon_{abcd} F^{cd}, \; a,b,..= 6,..,9
\ee
where
\be
\label{instanton-susy}
\Gamma_{6789}\epsilon = \epsilon
\ee
These are the instanton solutions of euclidean SYM$_4$. We assume that the 
instanton number is positive and equal to $Q_1$.
These solutions are characterised by moduli whose variation 
does not change the action of SYM$_6$. Promoting these moduli to slowly 
varying functions of $x_5,t$ we obtain stringy
solitons of the 5-brane theory. In order to identify 
these solitons as $D_1$ branes we have to show that the instanton number
density is a source of the Ramond 2-form $C^{(2)}$.

To do this consider the Chern-Simons terms of the world volume theory of the $D_1$ branes, \cite{Pol-tasi96}.
\be 
\label{brane-within-brane}
\mu_5 \int d^6 x [C^{(2)} \wedge F \wedge  F]
\ee
which shows that non-zero values of $F_{67}, F_{89}$ can act
as a source term for $C^{(2)}_{05}$. The latter corresponds to
a D1-brane stretching in the 5 direction. 

We can also verify the mass of the $D_1$ branes by simply 
evaluating the instanton action, and it turns out to be 
\be
\label{mass-bps-YM}
M=\frac{1}{g_s^2} (a_1g_sQ_1)
\ee
where 
\be
\nonumber
a_{1}=\frac{R}{\alpha'} \nonumber \\
\ee

This is a beautiful realization of a brane as a soliton bound 
within another brane. The motions of these $D_1$ branes iside the 5-branes 
represents the low lying collective modes of the $D_1-D_5$ system.
$R >> \sqrt{\alpha'}$ once more implies that we neglect the winding modes of the soliton strings. The KK modes of the YM$_6$ are also neglected. 

The technically difficult part here is that the moduli space of
instantons on the 4-torus is not a well known mathematical object. For
example the ADHM construction \cite{ADHM-ref} is valid for $R^4$ and
not $T^4$. There is a possibility that the ADHM construction for this
case, in the limit of small instanton size involves the 2 sets of
D-terms that we discussed in the last section. The advertised moduli
space ${\cal M}$, of instantons on $T^4$, is the Hilbert Scheme of the
symmetric product $(\tilde{T}^4)^{Q_1 Q_5}/S(Q_1Q_5)$. ($\tilde{T}^4$
can be different from the compactification torus $T^4$.)

One can give physical arguements to support at least the topological
aspect of the above result 
\cite{Vaf-Wit94,Vaf95-instanton,Vaf95-gas}. One uses the fact 
that the configuration
of $D_1-D_5$, we are working with, is U-dual to a fundamental
string with winding modes. The BPS states of this fundamental string
(that is, states with either purely left moving or right moving
oscillators) maps to the ground states of the $D_1-D_5$ system which is given
by the dimension of the cohomology of ${\cal M}$.

Our attitude will be to consider the sigma model on ${\cal M}$, as a
resolution of the sigma model on the orbifold $(\tilde{T}^4)^{Q_1
  Q_5}/S(Q_1Q_5)$. It so turns out that ${\cal M}$ is a hyper-kahler
manifold and hence one can define a $N=(4,4)$ SCFT. We will explicitly
construct the $N=(4,4)$ orbifold SCFT and discuss its blowing up
modes, which turn out to be 4 marginal operators of the SCFT. 

Before we do that we would like to discuss the validity of our considerations
in the strong coupling region where $g_sQ_1Q_5\gg 1$.

First we note that (\ref{instanton-equation}) is derived as a
condition from supersymmtry and is independent of the coupling
constant. These instantons also do not receive any stringy corrections
\cite{sei-wit-ncgeom}. After this we used the collective coordinate
method to arrive at the long wavelength approximation. However in the
standard procedure we have to assume that $g_s$ is small and hence we
can neglect the interactions of the collective coordinates with the
other Yang-Mills quanta. However it can be shown that
the moduli space and the corresponding sigma model does not receive
any corrections in the string coupling. This is basically because the
hypermultiplet moduli space does not get renormalized by the
interactions \cite {seiberg,maldacena}. This fact is crucial because
it says that the SCFT that we found at weak coupling is valid at
strong coupling and hence we can use it to make comparisons with
supergravity calculations.

\section{\bf The ${\cal N}=4$ Super Conformal Algebra}
We now discuss the super conformal algebra generated by the holomorphic 
stress tensor 
$T(z)$, a doublet of supersymmetry generators $G^a(z), G^{b\dagger}(z)$ and  $J^i(z)$ the $SU(2)$ $R$ local symmetry. There are also corresponding anti-holomorphic generators $\widetilde J(\bar z), \widetilde G(\bar z)$
and $\widetilde T(\bar z)$.
\bea
\label{scft_algebra}
T(z)T(w) &=& \frac{\del T(w)}{z-w} + \frac{2 T(w)}{(z-w)^2} +
\frac{c}{2 (z-w)^4},  \\  \nonumber
G^a(z)G^{b\dagger }(w) &=& 
\frac{2 T(w)\delta_{ab}}{z-w} + \frac{2 \bar{\sigma}^i_{ab} \del J^i}
{z-w} + \frac{ 4 \bar{\sigma}^i_{ab} J^i}{(z-w)^2} + 
\frac{2c\delta_{ab}}{3(z-w)^3}, \\
\nonumber
J^i(z) J^j(w) &=& \frac{i\epsilon^{ijk} J^k}{z-w} + \frac{c}{12 (z-w)^2}
, \\  \nonumber
T(z)G^a(w) &=& \frac{\del G^a (w)}{z-w} + \frac{3 G^a (z)}{2 (z-w)^2},
\\   \nonumber
T(z) G^{a\dagger }(w) &=& \frac{\del G^{a\dagger} (w)}{z-w} + 
\frac{3 G^{a\dagger} (z)}{2 (z-w)^2}, \\   \nonumber
T(z) J^i(w) &=& \frac{\del J^i (w)}{z-w} + \frac{J^i}{(z-w)^2}, \\
\nonumber
J^i(z) G^a (w) &=& \frac{G^b(z) (\sigma^i)^{ba}}{2 (z-w)}, \\
\nonumber
J^i(z) G^{a\dagger}(w) &=& -\frac{(\sigma^i)^{ab} G^{b\dagger
}(w)}{2(z-w)} 
\eea
The global R-parity group, is given by the zero modes of the currents $J^i(z)$
and $\widetilde J(\bar z)$. It is denoted by
$SU(2)_R \times \widetilde{SU(2)}_R$, and it is an outer automorphism 
of the N=(4,4) current algebra. The N=(4,4) SCFT admits another global $SO(4)$
symmetry which we shall discuss subsequently.

\subsection{The supergroup $SU(1,1|2)$}
We now discuss the zero mode part of the current algebra of 
the previous section. This is the Lie super-algebra $SU(1,1|2)$ generated 
by the global charges: $L_{\pm,0} ,J^{(1),(2),(3)}_R$ and  
$G^a_{1/2,-1/2}$. The
global charges of the supersymmetry currents $G^a(z)$ are
in the Neveu-Schwarz sector.
\bea
[L_0, L_{\pm}] = \mp L_{\pm} \;\;&\;&\;\; [L_{1} , L_{-1}] = 2L_{0} \\ \nonumber
\{ G^a_{1/2} , G^{b\dagger}_{-1/2} \} &=& 2\delta^{ab}L_0 + 2
\sigma^i_{ab} J^{(i)}_{R} \\  \nonumber
\{ G^a_{-1/2} , G^{b\dagger}_{1/2} \} &=& 2\delta^{ab}L_0 - 2
\sigma^i_{ab} J^{(i)}_{R} \\  \nonumber
[J^{(i)}_R, J^{(j)}_R] &=& i\epsilon^{ijk}J^{(k)}_R \\ \nonumber
[L_0, G^a_{\pm 1/2}] = \mp\frac{1}{2} G^a_{\pm 1/2} \;\;&\;&\;\;[L_0, G^{a\dagger}_{\pm 1/2}] = \mp\frac{1}{2} G^{a\dagger}_{\pm 1/2} 
\\ \nonumber
[L_+ , G^a_{1/2}] = 0 \;\;&\;&\;\; [L_- , G^a_{-1/2}] = 0 \\ \nonumber
[L_- , G^a_{1/2}] = -G^a_{-1/2} \;\;&\;&\;\; [L_+ , G^a_{-1/2}] = G^a_{1/2} \\ \nonumber
[L_+ , G^{a\dagger}_{1/2}] = 0 \;\;&\;&\;\; [L_- , G^{a\dagger}_{1/2}] = 0 \\ \nonumber
[L_- , G^{a\dagger}_{1/2}] = -G^a_{-1/2} \;\;&\;&\;\; [L_+ , G^{a\dagger}_{-1/2}] = G^a_{1/2} \\ \nonumber
[J^{(i)}_R, G^a_{\pm 1/2} ] = \frac{1}{2} G^{b}_{\pm 1/2} (\sigma^i)^{ba}
\;\;&\;&\;\; [J^{(i)}_R, G^{a\dagger}_{\pm 1/2} ] = -\frac{1}{2}
 (\sigma^i)^{ba}G^{b\dagger}_{\pm 1/2} \\  \nonumber
\eea
The anti-holomorphic sector leads to an identical algebra so that the global Lie super-algebra is $SU(1,1|2)\times
SU(1,1|2)$. 

One can clearly see that $SU(1,1|2)$ has 8 real supercharges and hence
we have a total of 16 real SUSYS, whereas the supergravity background,
with zero KK momentun along the circle $x_5$, has only 8 real SUSYS!

\subsection{Maldacena Duality: Geometry Dual to SCFT}
The puzzle of the doubling of the SUSYS is resolved by the remarkable discovery of Maldacena that 
the geometry dual to the SCFT is infact not the asymptotically flat space which supports 8 SUSYS, but $AdS_3\times S^3\times \tilde{T}^4$. The duality conjecture as far as the symmetries are concerned states that the $SU(1,1|2)\times SU(1,1|2)$ symmetry of the near
horizon geometry is matched with the global part of the ${\cal
N}=(4,4)$ SCFT on ${\cal M}$ together with the
identification of the $SO(4)_I$ algebra of $T^4$ and $\tilde{T}^4$.
\\

\begin{tabular}{ll}
Symmetries of the Bulk & Symmetries of SCFT \\
\hline
 & \\
(a) Isometries of $AdS_3$ 
&The global part of the Virasoro group \\
 $SO(2,2)\simeq SL(2, R) \times \widetilde{SL(2,R)}$ &
$SL(2,R)\times \widetilde{SL(2,R)}$ \\
 & \\
(b) Isometries of $S^3$ &
R-symmetry of the SCFT \\
$SO(4)_E\simeq SU(2)\times SU(2)$ &
$SU(2)_R\times\widetilde{SU(2)}_R$ \\
 & \\
(c) Sixteen near horizon symmetries & Global supercharges  of 
${\cal N}= (4,4) $ SCFT \\
& \\
(d) $SO(4)_I$ of $T^4$ & $SO(4)_I$ of $\tilde{T}^4$ \\
\hline
\end{tabular} \\

\gap

With this we conclude the general discussion of matching symmetries of
the SCFT and the $AdS_3\times S^3\times \tilde{T}^4$ geometry that is
dual to it. In the following section we discuss a specific
representation which has further symmetries that enable us to match
operators and fields on both sides of the dual description.

\section{\bf ${\cal N}=(4,4)$ SCFT on the
  orbifold $(\tilde{T}^4)^{Q_1 Q_5}/S(Q_1 Q_5)$} 

The ${\cal N}=4$
superconformal algebra with $c=6Q_1Q_5$ can be constructed out of
$Q_1Q_5$ copies of $c=6$,\, ${\cal N}=4$ superconformal algebra on
$\tilde{T}^4$. The discussion in this and subsequent sections is mainly taken from
\cite {Dav-Man-Wad99,Dav-thesis}.

The Lagrangian is given by
\be
\label{free}
S = \frac{1}{2} \int d^2 z\; \left[\del
x^i_A \bar\del x_{i,A} + 
\psi_A^i(z) \bar\del \psi^i_A(z) + 
\widetilde\psi^i_A(\bar z) \del \widetilde \psi^i_A(\bar z) 
 \right]
\ee
Here $i$ runs over the $\widetilde{ T^4}$ coordinates
1,2,3,4 (we have renamed the internal coordinates) and $A=1,2,\ldots,Q_1Q_5$ labels various copies
of the four-torus. The symmetric group $S(Q_1Q_5)$
acts by permuting the copy indices.

Let us introduce some definitions. The complex bosons $X$ and the fermions $\Psi$ are defined as:
\bea
\label{defn}
X_A(z) &=& (X^1_A(z), X^2_A(z)) = \sqrt{1/2} (x^1_A(z) + i x^2_A(z),
x^3_A(z) + i x^4_A(z)),   \\  \nonumber
\Psi_A (z) &=& (\Psi^1_A(z), \Psi^2_A(z)) = \sqrt{1/2} (\psi^1_A(z) +
i\psi^2_A(z), \psi^3_A(z) + i\psi^4_A(z)) \\   \nonumber
X_A^\dagger (z) &=& 
\left(
\begin{array}{c}
X_A^{1\dagger} (z) \\
X_A^{2\dagger} (z)
\end{array}
\right)   = \sqrt{\frac{1}{2}}
\left(
\begin{array}{c}
x^1_A(z)-ix^2_A (z)\\
x^2_A(z)-ix^2_A(z)
\end{array}
\right) \\   \nonumber
\Psi_A^\dagger (z) &=&
\left(
\begin{array}{c}
\Psi_A^{1\dagger} (z) \\
\Psi^{2_A \dagger} (z)
\end{array}
\right)     =\sqrt{\frac{1}{2}} 
\left(  
\begin{array}{c}
\psi^1_A(z) - i\psi^2_A(z)\\
\psi^3_A(z) - i\psi^4_A(z)
\end{array}
\right)
\eea
In terms of these we can write the generators of the SCFT,
\bea
T(z) &=& \del X_A (z)\del X^\dagger_A(z) 
+ \frac{1}{2}\Psi_A (z)\del \Psi^{\dagger}_A (z) 
- \frac{1}{2}\del\Psi_A (z) \Psi^{\dagger}_A (z) 
\\   \nonumber
G^a(z) &=&
\left(
\begin{array}{c}
G^1(z)\\
G^2(z)
\end{array}
\right) =
\sqrt{2} \left(
\begin{array}{c}
\Psi^1_A (z) \\
\Psi^2_A (z) \end{array}  \right)  \del X^2_A (z)  +
\sqrt{2} \left( \begin{array}{c}
-\Psi^{2\dagger}_A (z) \\
\Psi^{1\dagger}_A (z)  \end{array}  \right)  \del X^1_A (z) 
\\    \nonumber
J^i_R(z) &=& \frac{1}{2} \Psi_A(z)\sigma^i\Psi^\dagger_A (z)\\
\nonumber
\eea
The charges corresponding to the R-parity current are:
\be
J^i_R = \frac{1}{2}\int\frac{dz}{2\pi i} 
\Psi_A(z)\sigma^i\Psi^\dagger_A(z)
\ee
In the above the summation over $A$ which runs from $1$ to $Q_1Q_5$ is
implied.

\subsection{The $SO(4)$ Global Symmetry of the SCFT}
We now discuss global symmetries that are particular to the free field
representation that we have discussed above.
There are $2$
global $SU(2)$ symmetries which correspond to the $SO(4)$ rotations of
the $4$ bosons $x^i$. The corresponding charges are given by 
\bea
I_1^i &=& 
\frac{1}{4}\int\frac{dz}{2\pi i} X_A \sigma^i \del X_A^\dagger 
-\frac{1}{4}\int\frac{dz}{2\pi i} \del X_A \sigma^i X_A^\dagger 
+ \frac{1}{2}\int\frac{dz}{2\pi i}
\Phi_A \sigma^i \Phi_A^\dagger  \\  \nonumber
I_2^i &=& 
\frac{1}{4} 
\int\frac{dz}{2\pi i}{\cal X}_A\sigma^i\del{\cal X}_A^\dagger
-\frac{1}{4} 
\int\frac{dz}{2\pi i}\del{\cal X}_A\sigma^i{\cal X}_A^\dagger
\eea
Here 
\bea
{\cal X }_A = (X^1_A, -X^{2\dagger}_A) \;&\;&\;\;
{\cal X}^\dagger  =
\left(
\begin{array}{c}
X^{1\dagger}_A \\
-X^2_A 
\end{array}  \right) \nonumber \\
\Phi_A = (\Psi^1_A, \Psi^{2\dagger}_A ) \;&\;&\;\;
\Phi_A^\dagger =
\left(
\begin{array}{c}
\Psi^{1\dagger}_A  \\
\Psi^2_A 
\end{array} \right).
\eea
These charges generate the $SU(2)\times SU(2)$ algebra:
\bea
[I_1^i, I_1^j] = i\epsilon^{ijk} I_1^k \;&\;&\;\;
[I_2^i, I_2^j] = i\epsilon^{ijk} I_2^k 
\\  \nonumber
[I_1^i, J_2^j] &=&0
\eea
The new global charges have the following commutation relations 
with the local currents, 
\bea
\label{so4_on_g}
[I_1^i, G^a(z)] =0 \;&\;&\;\;
[I_1^i, G^{a\dagger}(z)] =0 \\ \nonumber 
[I_1^i, T(z)] =0 \;&\;&\;\;
[I_1^i, J(z)]=0 \\ \nonumber 
[I_2^i, {\cal G}^a(z)] = 
\frac{1}{2}{\cal G}^{b} (z)\sigma^i_{ba} \;&\;&\;\;
[I_2^i, {\cal G}^{a\dagger} (z) ]
= - \frac{1}{2}\sigma^i_{ab}{\cal G}^{b\dagger}(z) \\ \nonumber
[I_2^i, T(z)] =0  \;&\;&\;\;
[I_2^i, J(z)]=0
\eea
where
\bea
{\cal G} = ( G^1, G^{2\dagger}) \;&\;&\;\;
{\cal G}^\dagger=
\left(
\begin{array}{c}
G^{1\dagger} \\
G^2
\end{array}
\right)
\eea

The charges $I_1, I_2$ constructed
above  generate $SO(4)$
transformations only on  the {\em holomorphic} bosons $X_A(z)$. 
Similarly, we can construct charges
$\widetilde{I_1}, \widetilde{I_2}$ which generate $SO(4)$
transformations only on  the {\em antiholomorphic} 
bosons $\widetilde{X_A}(\bar z)$. 
Normally one would expect these
charges to give rise to a global $SO(4)_{hol}
\times SO(4)_{antihol}$ symmetry. 
Howver a boson field is a sum of a holomorphic and anti-holomorphic part, 
and hence it has a well defined transformation property only under the charges
\bea
J_I= I_1  + \widetilde{I}_1 \;\;&\;&\;
\widetilde{J}_I = I_2 + \widetilde{I}_2
\eea
These generate the $SO(4)_I =
SU(2)_I\times \widetilde{SU(2)}_I$, 
and fall into representations of the
${\cal{N}}=(4,4)$ algebra (as can 
be proved by using the commutation relations \eq{boson}
of the $I$'s). The bosons $X(z,\bar{z})$
transform as $(\bf 2 , \bf 2)$ under $SU(2)_I\times \widetilde{SU(2)}_I$. 

The following commutations relation show that 
the bosons transform as $(\bf 2, \bf 2)$ under $SU(2)_{I_1}\times
SU(2)_{I_2}$
\bea
\label{boson}
[I_1^i, X^a_A] = \frac{1}{2} X^b_A\sigma^i_{ba} \;&\;&\;\;
[I_1^i, X^{a\dagger}_A] = -\frac{1}{2}\sigma^i_{ab}X^{b\dagger}_A 
\\   \nonumber
[I_2^i, {\cal X}^a_A ] =
\frac{1}{2} {\cal X}^b_A \sigma^i_{ba}  \;&\;&\;\;
[I_2^i, {\cal X}^{a\dagger}_A] =
-\frac{1}{2}\sigma^i_{ab}{\cal X}^{b\dagger}_A 
\eea
The fermions transform as $(\bf 2, \bf 1)$ under $SU(2)_{I_1}\times 
SU(2)_{I_2}$ as can be seen from the commutations relations
given below.
\bea
[I_1^i, \Phi^a_A] = \frac{1}{2}\Phi^b_A \sigma^i_{ba} \;&\;&\;\;
[I_1^i, \Phi^{a\dagger}_A] =
-\frac{1}{2}\sigma^i_{ab}\Phi^{b\dagger}_A  \\   \nonumber
[I_2^i, \Psi^a] =0  \;&\;&\;\;
[I_2^i, \bar{\Psi}^a]=0
\eea
 
\section{\bf $SU(1,1|2)$ Classification of States of the SCFT}
The $SU(1,1|2)$ algebra has 2 sub-algebras: the global Virasoro algebra
and the $SU(2)_R$ algebra. Their representations are labeled by the conformal 
weight h and the $SU(2)_R$ spin j. The highest weight states
$ |\mbox{hw}\rangle = 
|h,{\bf j}_R,j_R^3 =j_R \rangle $ are defined by,
\bea
L_1 |\mbox{hw}\rangle = 0 \;\;\; 
L_0 |\mbox{hw}\rangle = h|\mbox{hw}\rangle \\ \nonumber
J^{(+)}_{R}|\mbox{hw}\rangle =0  \;\;\; 
J_R^{(3)}|\mbox{hw}\rangle = j_R|\mbox{hw}\rangle\\  \nonumber
G_{1/2}^a|\mbox{hw}\rangle =0 \;\;\; G_{1/2}^{a\dagger}
|\mbox{hw}\rangle =0
\eea
$J^+_R$ is the raising operator for spin $j_R^3$

Amongst these highest weight states those for which $h=j$ satisfy additional
conditions 
$
G^{2\dagger }_{-1/2}|\mbox{hw}\rangle =0 ,\;\;\;
G^1_{-1/2}|\mbox{hw}\rangle =0
$. 
These states are called{ \it{chiral primaries}}. From the chiral primaries one can 
generate multiplets by the action of the operators 
$G^{1\dagger }_{-1/2}$ and $G^2_{-1/2}$. These multiplets are 
called {\it{short multiplets}}. The $h=j$ short multiplet is given in the following table:
\bea
\label{short}
\begin{array}{cccc}
\mbox{States} & j & L_0 & \mbox{Degeneracy} \\
|\mbox{hw}\rangle_{S} & h & h& 2h+1 \\
G^{1\dagger }_{-1/2}|\mbox{hw}\rangle_{S}, 
G^2_{-1/2}|\mbox{hw}\rangle_{S} &h-1/2& h+1/2 & 2h + 2h = 4h \\
G^{1\dagger }_{-1/2} G^2_{-1/2}|\mbox{hw}\rangle_{S} & h-1& h+1& 2h-1
\end{array}
\eea
The short multiplets are usually denoted by the degeneracy of the $h=j$ state.

Since we have $SU(1,1|2)$ also comming from the anti-holomorphic sector the 
global super-algebra is $SU(1,1|2)\times SU(1,1|2)$ and its short multiplets will 
be denoted by $(\bf{2h +1}, \bf{2h'+1})_S$. The top component of
the short multiplet are the states belonging to  the last row in
\eq{short}. These states are annihilated by {\it{all}} the super-charges.

\subsection{The Chiral Primaries and Marginal Operators of the Untwisted Sector}
The short multiplet $(\bf{2}, \bf{2})_S$ is special, it
terminates at the middle row of \eq{short}. For this case, the top
component is the middle row. These states have $h=\bar{h}=1$
and transform as $(\bf{1}, \bf{1})$ of $SU(2)_R\times
\widetilde{SU(2)}_R$. There are $4$ such states for each $(\bf{2},
\bf{2})_S$. Hence these give rise to 16 marginal deformations of the SCFT.
We shall see that there are 4 more marginal operators which come from the $Z_2$
twisted sector of the SCFT.

The 16 marginal operators of the untwisted sector arise as short
multiplets belonging to the 4 chiral primaries $(\bf{2},\bf{2})_S$,
\bea 
\label{chiral}
\Psi^1_A(z) \widetilde{\Psi}^1_A(\bar{z})  \;&\;&\;  
\Psi^1_A(z)\widetilde{\Psi}^{2\dagger}_A(\bar z) \\   \nonumber
\Psi^{2\dagger}_A(z)\widetilde{\Psi}^1_A(\bar z )  \;&\;&\;
\Psi^{2\dagger}_A(z)\widetilde{\Psi}^{2\dagger}_A(\bar z) 
\eea
where summation over $A$ is implied.These four operators have conformal
dimension $(h, \bar{h})=(1/2, 1/2)$ and
$(j_R^3, \widetilde{j}_R^3)= (1/2, 1/2)$ under
the R-symmetry  $SU(2)_R\times \widetilde{SU(2)}_R$.
These are the relevant operators of the SCFT.

The short multiplets corresponding to each of the above chiral
primaries can be constructed following the table in the previous
section. Each such multiplet leads to 4 marginal operators with
conformal weights $(1,1)$ and
transform as $(\bf 1, \bf 1 )$ under $SU(2)_R\times
\widetilde{SU(2)}_R$. These operators can be derived from the
pole terms of the operator product expansion of the chiral primaries
with the SUSY currents. The result agrees with the expectation that
the $16$ top components of the $4 (\bf 2, \bf 2)_S$ short multiplets
are $\del x_A^i \bar{\del} x_A^j$. These top components
can be added to the SCFT as
perturbations without violating the ${\cal N}=(4,4)$ supersymmetry.
 
It is clear that these operators 
can be classified using the global $SO(4)_I$ symmetry
of the SCFT. The four torus $\tilde{T}^4$ actually breaks this
symmetry but we assume that the target space is $R^4$ for the
classification of states. We have the following table of various quantum numbers,
\bea
\label{untwist_operator}
\begin{array}{lccc}
\mbox{Operator}&SU(2)_I\times \widetilde{SU(2)}_I&
SU(2)_R\times\widetilde{SU(2)}_R& (h, \bar{h}) \\
\del x^{ \{ i }_A(z) \bar{\del}x^{ j\} }_A (\bar z) -
\frac{1}{4}\delta^{ij}
\del x^k_A(z) \bar{\del}x^k_A (\bar z) &(\bf 3, \bf 3) & (\bf 1,\bf 1) 
& (1, 1) \\  
\del x^{[i}_A(z) \bar{\del}x^{j]}_A (\bar z) & (\bf 3, \bf 1) +
(\bf 1, \bf 3) & (\bf 1, \bf 1)& (1,1) \\ 
\del x^i_A(z) \bar{\del}x^i_A (\bar z) &(\bf 1, \bf 1)& (\bf 1, \bf 1)
& (1,1)
\end{array}
\eea

\subsection{Chiral Primaries and Marginal Operators of the $Z_2$ Twist Sector }
The orbifold SCFT has twisted sectors corresponding to conjugacy classes of the 
symmetric group $S(Q_1 Q_5)$. These classes are labeled by cyclic groups 
of various lengths. If $n$ is the length of the cycle and $N_n$ is it's multiplicity 
then we have the basic equation,
\be
\sum nN_n = Q_1 Q_5
\ee

The simplest conjugacy class consists of 1 cycle of length 2, and $Q_1 Q_5 -2$
cycles of length 1. An example of an element of this class is 
\be
\label{group_element}
(X_1\rightarrow X_2, \; X_2\rightarrow X_1), 
\; X_A\rightarrow X_A, A=1, \ldots ,Q_1 Q_5 -2 
\ee
Clearly the group action has a fixed point at $X_1=X_2$. The linear combination that carries a representation of $Z_2$ is
\be
X_{cm} = X_1+ X_2 \;\; \mbox{and}\;\; \phi = X_1-X_2
\ee
Under the group action \eq{group_element}\ 
$X_{cm}$ is invariant and $\phi
\rightarrow -\phi$. Thus the singularity is {\em locally} of the type
$R^4/Z_2$ or equivalently $C^2/Z_2$. The bosonic twist operators for this orbifold singularity
are given by following OPE's \cite{DixFriMarShe}
\bea
\del \phi^1 (z) \sigma^1(w, \bar{w} ) = 
\frac{ \tau^1(w, \bar w ) }{ (z-w)^{1/2} }  \; &\;& \;
\del {\phi}^{1\dagger} (z) \sigma^1(w, \bar{w} ) = 
\frac{ \tau'^1(w, \bar w ) }{ (z-w)^{1/2} }  \\   \nonumber
\del \phi^2 (z) \sigma^2(w, \bar{w} ) =
\frac{ \tau^2(w, \bar w ) }{ (z-w)^{1/2} }  \; &\;& \;
\del {\phi}^{2\dagger} (z) \sigma^2(w, \bar{w} ) =
\frac{ \tau'^2(w, \bar w ) }{ (z-w)^{1/2} }  \\  \nonumber
\bar{\del} \widetilde{\phi}^1 (\bar{z}) \sigma^1(w, \bar{w} ) = 
\frac{ \widetilde\tau'^1(w, \bar w ) }{ (\bar z-\bar w)^{1/2} }  \; &\;&
\;
\bar{\del} \widetilde{\phi}^{1\dagger} (\bar z) \sigma^1(w, \bar{w} ) = 
\frac{ \widetilde\tau^1(w \bar w ) }{ (\bar z-\bar w)^{1/2} } 
\\   \nonumber
\bar{\del} \widetilde{\phi}^2 (\bar z) \sigma^2(w, \bar{w} ) =
\frac{ \widetilde\tau'^2(w, \bar w ) }{ (\bar z-\bar w)^{1/2} }  \; &\;&
\;
\bar{\del} \widetilde{\phi}^{2\dagger} (\bar z ) \sigma^2(w, \bar{w} ) =
\frac{ \widetilde\tau^2(w, \bar w ) }{ (\bar z-\bar w)^{1/2} }  
\eea
The $\tau$'s are excited twist operators.
The fermionic twists are constructed from bosonized currents defined
by
\bea
\chi^1(z) = e^{iH^1(z)} \; &\;& \; \chi^{1\dagger}(z) = e^{-iH^1(z)} \\
\nonumber
\chi^2(z) = e^{iH^2(z)} \; &\;& \; \chi^{2\dagger}(z) = e^{-iH^2(z)} \\
\nonumber
\eea
Where the $\chi$'s, defined as
$\Psi_1 - \Psi_2$, are the superpartners of the bosons $\phi$.

>From the above the supersymmetric twist fields which act
both on fermions and bosons are:
\bea
\label{defSigma}
\Sigma^{(\frac{1}{2}, \, \frac{1}{2})}_{(12)} = \sigma^1(z,\bar z)
\sigma^2 (z,\bar z)
e^{iH^1(z)/2} e^{-iH^2(z)/2} 
e^{i\widetilde{H}^1(\bar z )/2} e^{-i\widetilde{H}^2(\bar z )/2}
\\  \nonumber
\Sigma^{(\frac{1}{2},\,  -\frac{1}{2})}_{(12)} = \sigma^1(z,\bar z)
\sigma^2 (z,\bar z)
e^{iH^1(z)/2} e^{-iH^2(z)/2} 
e^{- i\widetilde{H}^1(\bar z )/2} e^{i\widetilde{H}^2(\bar z )/2} 
\\  \nonumber
\Sigma^{(-\frac{1}{2}, \,  \frac{1}{2})}_{(12)} = \sigma^1(z,\bar z)
\sigma^2 (z,\bar z)
e^{-iH^1(z)/2} e^{+iH^2(z)/2} 
e^{i\widetilde{H}^1(\bar z )/2} e^{-i\widetilde{H}^2(\bar z )/2} 
\\   \nonumber
\Sigma^{(-\frac{1}{2}, \,  -\frac{1}{2})}_{(12)} = 
\sigma^1(z,\bar z) \sigma^2 (z,\bar z)
e^{-iH^1(z)/2} e^{+iH^2(z)/2} 
e^{-i\widetilde{H}^1(\bar z )/2} e^{+i\widetilde{H}^2(\bar z )/2} 
\eea\\  \nonumber
The subscript $(12)$ refers to the fact that these twist operators were
constructed for the representative
group element \eq{group_element}\ which exchanges the $1$ and
$2$ labels of the coordinates of $\widetilde{T}^4$. 
The superscript stands
for the $(j^3_R, \widetilde{j}^3_R)$ quantum numbers.
The full twist operators for the $Z_2$
conjugacy class are obtained by summing over 
$Z_2$ twist operators for all representative elements of this class.
For $(j^3_R, \widetilde{j}^3_R)=(1/2,1/2)$, we have
\be
\Sigma^{(\frac{1}{2}, \,  \frac{1}{2})} =
\sum_{i=1}^{Q_1Q_5} \sum_{j=1, j\neq i}^{Q_1Q_5}
\Sigma^{(\frac{1}{2}, \,  \frac{1}{2})}_{(ij)}
\ee
A similar construction holds for the 3 other operators. 
The conformal dimensions of these operators is
$(1/2,1/2)$. They transform as $(\bf 2 , \bf 2)$ under the $SU(2)_R
\times \widetilde{SU(2)}_R$ symmetry of the SCFT. They belong to the
bottom component of the short multiplet $(\bf 2, \bf 2)_S$. The
operator $\Sigma^{(\frac{1}{2}, \, \frac{1}{2})}$ is a chiral primary.
As before the $4$ top components of this short multiplet,
which we denote by
\bea
T^{(\frac{1}{2}, \, \frac{1}{2})}, \;\; T^{(\frac{1}{2}, \,
-\frac{1}{2})} \\ \nonumber
T^{(-\frac{1}{2}, \, \frac{1}{2})}, \;\;
T^{(-\frac{1}{2}, \, -\frac{1}{2})} 
\eea 
are given  by the leading pole in the following OPE's respectively
\bea 
\label{raising}
G^2(z)\widetilde{G}^2(\bar z)\Sigma^
{(\frac{1}{2}, \,  \frac{1}{2})} (w, \bar w), \;\;
G^2(z) \widetilde{G}^{1\dagger}(\bar z)\Sigma^{(\frac{1}{2}, \,  
\frac{1}{2})} (w, \bar w),  \\  \nonumber
 G^{1\dagger}(z) \widetilde{G}^2(\bar z)
 \Sigma^{(\frac{1}{2}, \, \frac{1}{2})} (w, \bar w), \;\;
G^{1\dagger}(z)\widetilde{G}^{1\dagger}
(\bar z)\Sigma^{(\frac{1}{2}, \,\frac{1}{2})} (w, \bar w)
\eea

The table of the quantum numbers of
these operators is given below.
\bea
\begin{array}{lccc}
\label{twist_operator}
\mbox{Operator} & (j^3, \widetilde{j}^3)_I & 
SU(2)_R\times \widetilde{SU(2)}_R &  (h, \bar{h}) \\
{\cal T}^1_{(1)}=T^{(\frac{1}{2}, \,  \frac{1}{2})} & 
( 0, 1) & (\bf 1, \bf 1) & (1,1) \\
{\cal T}^1_{(0)}= T^{(\frac{1}{2}, 
\,  -\frac{1}{2})}+ T^{(-\frac{1}{2}, \,  \frac{1}{2})} &  
(0,0) & (\bf 1, \bf 1) & (1,1) \\
{\cal T}^1_{(-1)}= T^{(-\frac{1}{2}, \,  -\frac{1}{2})} 
& (0,-1) & (\bf 1, \bf 1) & (1,1) \\
{\cal T}^0= T^{(-\frac{1}{2}, \, -\frac{1}{2})} - T^{(-\frac{1}{2}, 
\,  -\frac{1}{2})}
& (0, 0) & (\bf 1, \bf 1) & (1,1) 
\end{array}  
\eea

The first three operators of the above table can be organized as a
$(\bf 1, \bf 3)$ under $SU(2)_I\times\widetilde{SU(2)}_I$. We will
denote these $3$ operators as ${\cal T}^1$. The last
operator transforms as a scalar $(\bf 1, \bf 1)$ under 
$SU(2)_I\times\widetilde{SU(2)}_I$ and is denoted by ${\cal T}^0$. 

These marginal operators are the $4$ blow up modes of the $R^4/Z_2$ singularity
\cite{CveDix} 
\footnote{Relevance of $Z_2$ twist operators
to the marginal deformations of the SCFT has earlier
been discussed in \cite{HasWad1},\,\cite{DijVerVer}}. 
Since these are top components of the short multiplet $(\bf 2 , \bf
2)_S$ they can be added to the free SCFT as perturbations without
violating the ${\cal N} = (4,4)$ supersymmetry of the SCFT.

In conclusion we have accounted for the 20 marginal operators of the
SCFT: 16 from the untwisted sector and 4 from the twisted sector.
Those from the twisted sector have a special significance because a
non-singular SCFT correspondes to turning on non-zero values for the
corresponding moduli. 

It is also possible to show \cite{cecotti} using ${\cal N} = (4,4)$ supersymmetry that 
the moduli space of the these 20 marginal operators is the coset space
\be
\frac{SO(4,5)}{SO(4)\times SO(5)}
\ee
Further, the number of marginal operators is  $4h_{11}$ where $h_{11}$
is the Hodge number and corresponds to
the number of chiral primaries of weight $(h, \bar {h} ) = (1/2, 1/2)$.

Later we shall see that turning on these moduli
also corresponds to a true rather than a marginal bound state of the
brane system.

\subsection{ Chiral Primaries of Higher Twisted Sectors}
Cyclic groups of length $k$ lead to twisted sectors characterised by the discrete group
$Z_k$. In the vicinity of the fixed points the orbifold has the structure
\be
R^4\times R^4/\omega \times R^4/\omega^2 \times \ldots \times 
R^4/\omega^{k-1}
\ee
The coordinate $\phi_m$ is twisted by the phase $\omega^{m}$ ( $m$ runs
from $1\ldots k $).
The dimension of the supersymmetric twist operator which twists the
coordinates by a phase $e^{2\pi i n /N}$ in $2$ complex dimensions is
$h(n,N)= n/N$ \cite{DixFriMarShe}. Hence the dimension of the twist operator
corresponding to the cyclic group of length $k$ 
is given by
\be
h = \sum_{i=1}^{k-1} h(i,k) = (k-1)/2
\ee 
A similar formula holds in the anti-holomorphic sector. 
There is a twist operator corresponding to every element in the conjugacy class
and by summing over all the elements we can construct a chiral 
primary operator  $\Sigma^{(k-1)/2}$. It has
conformal dimension 
$(h, \bar{h}) = ((k-1)/2, (k-1)/2)$ and $(j_R^3, \tilde{j}_R^3)
=((k-1)/2, (k-1)/2)$.
It belongs to the bottom component of the 
short multiplet $(\bf{k}, \bf{k})_S$.
The other components of the shortmultiplet $(\bf{k}, \bf{k})_S$ 
corresponding to the k-cycle twists can be generated by the
action of supersymmetry currents and the R-symmetry currents of the
${\cal N}=(4,4)$ theory.

As the largest cycle is of length $Q_1Q_5$, the
maximal dimension and angular momentum of the k-cycle twist operator is 
$((Q_1Q_5-1)/2, (Q_1Q_5-1)/2)$. This implies the important conclusion that 
the maximal value of the angular momentum of the corresponding state is $(Q_1Q_5-1)/2$.
This statement is called the {\it{stringy exclusion principle}} 
\cite{Mal-Str98}.

The chiral primaries with conformal weight $(h, \bar{h})$ of a ${\cal
  N}=(4,4)$ superconformal field theory on a manifold $K$ correspond
to the elements of the cohomology ${\cal H}_{2h\, 2\bar{h}}(K)$
\cite{Wit-Susy}. The chiral primaries are formed by the product
of the chiral primaries corresponding to the cohomology of the
diagonal $\tilde{T}^4$ denoted by $B^4$ (the sum of all copies of
$\tilde{T}^4$) and the various k-cycle chiral primaries. 

The cohomology of $B^4$ is constructed in terms of the complex fermions
defined in (\ref{defn}).
The elements of ${\cal H}_{11}(B^4)$ were already 
given in (\ref{chiral}). ${\cal H}_{22}(B^4)$ 
consists of the top form of $B^4$
\be
\Psi^1_A(z) \Psi^{2\dagger}_A(z)
\widetilde{\Psi}^1_A(\bar{z})  \widetilde{\Psi}^{2\dagger}_A(\bar z) 
\ee
where summation over all indices of $A$ is implied.

The chiral primaries of $B^4$ which correspond to
the elements of the cohomology ${\cal H}_{10}(B^4)$ are given by
\be
\sum_{A=1}^{Q_1Q_5}
\Psi^1_A(z) \;\;\; \mbox{and} \;\;\;
\sum_{A=1}^{Q_1Q_5}
\Psi^{2\dagger}_A(z)
\ee
and those that 
correspond to the
elements of the cohomology ${\cal H}_{21}(B^4)$  are 
\be
\Psi^1_A(z) \Psi^{2\dagger}_A(z)
\widetilde{\Psi}^1_A(\bar{z})  \;\; \mbox{and} \;\;
\Psi^1_A(z) \Psi^{2\dagger}_A(z)
\widetilde{\Psi}^{2\dagger}_A(\bar z) 
\ee
 
${\cal H}_{20}(B^4)$ has only one element which is given by
\be
\Psi^1_A(z) \Psi^{2\dagger}_A(z)
\ee
where summation over $A$ is implied. 

Once we know cohomology of $B^4$ the cohomology of ${\cal M}$ can be
easily constructed by combining the chiral the chiral primaries
$\Sigma^{k/2}(z, \bar{z})$ of the various twisted sectors. For details see
\cite{Dav-thesis}.
Below we present the answer for the set of chiral primaries with $h=\bar h$.
\bea
\label{eq.short}
&5 (\re 2 , \re 2 )_S + 6 \oplus_{\re m \geq \re 3 } (\re m , \re m
)_S \\ \nonumber
&+ 5((Q_1Q_5 + )1, (Q_1Q_1 + 1))_S + ((Q_1Q_5 + 2), (Q_1Q_1 + 2) )_S
\eea
In the above the maximaum value of $\re m = Q_1Q_5$. 

\subsection{Matching Chiral Primaries with States in Supergarvity}
We have already discussed the Maldacena duality conjecture in
section $10.2$. The global symmetries of the SCFT exactly match the
symmetries of $AdS_3\times S^3$. Further the radius of $S^3$ is
$\sqrt{\alpha'} (g_6^2 Q_1Q_5)^{1/4}$. Since this is a very large
number in the supergravity limit $g_sQ_1>>1 ,g_sQ_5>>1$, the masses of
the Kaluza-Klien modes on $S^3$ are very small and we expect these to
match the states of the short multiplets of the SCFT in the limit when
$Q_1$ and $Q_5$ are very large.
This indeed happens (except for short multiplets that correspond to non-propagating degrees of freedom) We refer the reader to the literature for details 
\cite {deB98, Dav-Man-Wad98, Mal-Moo-Str99}.

\section{The Supergravity Moduli and Correspondence with SCFT}
In this section we will match the SUGRA moduli in the near 
horizon geometry and marginal operators of the SCFT. 

Let us first discuss the moduli of supergravity.  It is known that the
moduli space of type IIB sugra compactified on a 4-torus consists of
25 scalars which parametrize the coset $SO(5,5)/(SO(5)\times SO(5))$.
These correspond to $10$ scalars $h_{ij}$ which arise from
compactification of the metric.  $i, j, k \ldots $ stands for the
directions of $T^4$, $6$ scalars $b_{ij}$ which arise from the
Neveu-Schwarz $B$-field, $6$ scalars $b'_{ij}$ from the Ramond-Ramond
$B'$-field, 3 scalars are the ten-dimensional dilaton $\phi_{10}$, the
Ramond-Ramond scalar $\chi$ and the Ramond-Ramond $4$-form $C_{6789}$.
In the near horizon geometry 5 of the above scalars become massive.
These correspond to $h_{ii}$, which is proportional to the volume of
the 4-torus, $b_{ij}^-$, the anti-self dual part of the Neveu-Schwarz
$B$ field and a certain linear combination of the RR 4-form and
scalar.  This moduli space corresponds to the coset
$SO(5,4)/(SO(5)\times SO(4))$.

Now since the above fields are massless we can use the isometries of $AdS_3$ to calculate the conformal dimensions $(h,\bar h)$.
The mass formula is given by \cite {Wit98-ads, Mal-Str98}
\be
h+\bar h = 1+ \sqrt {1+m^2}
\ee
Using this we note that the massless fields have $(h,\bar h)$=
$(1,1)$ and hence they belong to the top component of the short multiplet 
$5(\bf 2, \bf 2 )_S$. 

The quantum numbers of these massless states are summarized in the table below:
\bea
\label{sugra_fields}
\begin{array}{lccc}
\mbox{Field} & SU(2)_I\times \widetilde{SU(2)_I} 
& SU(2)_E\times \widetilde{SU(2)}_E &
\mbox{Mass} \\
h_{ij } -\frac{1}{4}
\delta_{ij} h_{kk} & (\bf 3, \bf 3) & (\bf 1, \bf 1) & 0 \\
b'_{ij} & (\bf 3, \bf 1) + (\bf 1, \bf 3) &(\bf 1, \bf 1) &  0 \\
\phi_6 &  (\bf 1, \bf 1) &(\bf 1, \bf 1) &  0 \\
a_1 \chi + a_2 C_{6789}&  (\bf 1, \bf 1) &(\bf 1, \bf 1) &  0 \\
b^+_{ij} &  (\bf 1, \bf 3) &(\bf 1, \bf 1) & 0 
\end{array}
\eea

In section 2.1 and 2.2 we have presented a table of the quantum numbers of the SCFT marginal operators. We would now like to match 
those marginal operators with the sugra fields we have obtained above.

We give the answer below and then justify it:
\bea
\begin{array}{llc}
\mbox{Operator} & \mbox{Field}  &  
SU(2)_I\times \widetilde{SU(2)}_I \\
\del x^{ \{ i }_A(z) \bar{\del}x^{ j\} }_A (\bar z) -1/4\delta^{ij}
\del x^{k}_A \bar{\del}x^k_A
& h_{ij} -1/4\delta_{ij} h_{kk} & (\bf 3, \bf 3 ) \\
\del x^{[i}_A(z) \bar{\del}x^{j]}_A (\bar z) 
& b'_{ij}
& (\bf 3, \bf 1 ) +
(\bf 1, \bf 3)  \\ 
\del x^i_A(z) \bar{\del}x^i_A (\bar z) 
& \phi
&( \bf 1, \bf 1 ) \\
{\cal T}^1 & b^+_{ij} & (\bf 1, \bf 3 ) \\
{\cal T}^0 & a_1\chi + a_2C_{6789} & (\bf 1, \bf 1 )
\end{array}
\eea

The representations ${\bf (1,3)}$ and ${\bf (1,1)}$ occur twice in the
above table and hence there may be an ambiguity in the identification
of these sugra fields with the SCFT operators. We resolve the
ambiguity with the help of the following arguement.  We have noted
that the operators ${\cal T}^1$ and ${\cal T}^0$ correspond to blowing
up modes of the orbifold CFT.  Turning these off would lead us to a
singular SCFT. This singularity has been related to the fact that in a
marginal bound state of the brane system one can have fragmentation
into sub-systems.  In the supergravity as we have explicitly seen
turning on the self-dual $b^+_{ij}$ Neveau-Schwarz field leads to a
stable bound state of the D1-D5 system. This is also true of the
modulus corresponding to $a_1\chi + a_2C_{6789}$. Solutions with these
moduli turned off correspond to marginal bound states. Hence we expect
that the blowing up modes of the SCFT correspond to the stabilizing
moduli.

\subsection{The Maximally Twisted Sector and Black Hole States}
The black hole is represented by a density matrix 
\be
\label{chp4:density-matrix}
\rho= \frac{1}{\Omega} \;\sum_{ \{i\} } |i\rangle\langle i|  
\ee
The states $|i\rangle$ 
belongs to the various twisted sectors of the orbifold theory.
The total value of $L_0$ and $\bar{L}_0$ satisfy the constraint 
\be
L_0=N_L \;\;\;\; \bar{L}_0=N_R
\ee
This corresponds to the fact that the general non-extremal black hole will have Kaluza-Klein 
excitations along both the directions on the $S^1$.

As we have seen before this information is sufficient to give the entropy of the degenerate state which satify the above constraint.
We will see that infact this contribution for large values of
the charges comes from the maximally twisted sector of the SCFT.
The maximally twisted sector of the orbifold CFT coresponds to the
longest cycle of the symmetric group $S(Q_1Q_5)$. It has a corresponding chiral primary $\sum^{(Q_1 Q_5-1)/2}$ and it's associated short multiplet. The presence of the twist field is equivalent to 
the following boundary condition on the bosonic fields
\be
\label{chp4:bc}
X_A (e^{2\pi i}z,e^{-2 \pi i} \bar{z})=X_{A+1} (z,\bar{z})
\ee
This implies that the momentum $n_L, n_R$ in the twisted sector is quantized in units of $1/(Q_1Q_5)$, and hence the monentum quantum number can go upto an integer multiple of $(Q_1Q_5)$.
Hence the contribution to the entropy from the twisted sector is 
\be
S(\mbox{maximally twisted})=2 \pi \sqrt{n_L} + 2 \pi \sqrt{n_R}
\ee
This equals the total entropy with the choice $n_L=Q_1Q_5N_L$
and $n_R=Q_1Q_5N_R$.

Hence we can identify the black hole micro-states with the states in the maximally twisted sector.

\section{Hawking Radiation}
In this section we discuss the derivation the Hawking process from the view point of string theory. First let us collect all the ingredients 
we have to formulate the process.
\begin{enumerate}
\item We have a complete description of the low lying excitations of
the D1-D5 system representated as a SCFT.
\item We have a description of the black hole microstates in terms of
fractionally moded oscillators that correspond to the chiral primary
$\sum^{(Q_1 Q_5-1)/2}$. 
\item Both the above statements about the microscopic theory are valid 
at large values of the effective coupling $g_s\sqrt {Q_1 Q_5}$ where the supergravity description is also valid.

If we model the black hole by a density matrix
\be
\rho= \frac{1}{\Omega}\sum_i | i \rangle \langle i |
\ee
where $| i \rangle$ represents the microstates then 
this sector of the full Hilbert space accounts
for the black hole entropy
\item We have a correspondence between the massless 
supergravity modes and the marginal operators of the SCFT.
\end{enumerate}.

The Hawking process corresponds to a transition from one black hole
state $| i \rangle$ to another $| f \rangle$ with the emission of a particle. The emitted particle
in principle corresponds to any state allowed by the symmetries.
However we shall restrict ourselves to the the massless emissions
which are a predominant decay mode.

The description of the above transition requires an interaction hamiltonian. In the absence of a Born-Infeld action we appeal to the symmetries of the problem. From the discussion of the classification and matching of the marginal operators of the SCFT and the massless fields of supergravity in the near horizon limit, we can formulate a simple interaction hamiltonian to first order which is consistent with the symmetries.
\be
S_{int} = \int d^2 z  \varphi_n|_B O_n(z, \bar z) 
\ee
In the above $\varphi_n|_B$ stands for the boundary value of the closed string (supergravity) mode, and $ O_n(z, \bar z)$ stands for the corresponding operator. We will assume that the normalization of the supergravity mode is such that in the bulk theory it leads to a standard
kinetic energy term. Using the above interaction one can calculate the S-matrix element relevant to the transition from one micro-state to another with the emission of a particle that couples to the micro-state.
\be
S_{if} = \langle f | H_{int} | i \rangle
\ee
Note that this S-matrix describes a transition from one pure state to another. 
If we use the density matrix description then we have to average over all the initial states to obtain the probability of absorption,
\be
\rm Prob_{abs}= \frac{1}{\Omega}\sum_i \sum_f |S_{if}|^2
\ee
where $\Omega$ is the number of initial microstates.

Let us now use the above principles to calculate the absorption cross-section and the Hawking rate corresponding to one of the 20 minimal scalars namely, $h_{ij}$ whose corresponding operator was found to be $\del x^i_A \bar \del x^j_A$. Both transform ${(\bf 3,\bf 3)}$ under $SO(4)_I$. The invariant interaction is given by
\be
\label{interaction} 
S_{\rm int} = \frac{\mu}{2}  \int d^2 z\; 
\left[ h_{ij} \del_z x^i_A \del_{\bar z} x^j_A \right]
\ee
The undetermined constant $\mu$ can be absorbed in the normalization
of the SCFT operator which inturn cannot be fixed within the frame of
the conformal field theory. We fix this normalization by matching the
2-point function of $\del x^i_A \bar \del x^j_A$ in the SCFT and the
2-point function of the internal graviton $h_{ij}$, in accordance with
the Ads/CFT conjecture. This matching implies (in our
conventions) $\mu=1$ \cite {Dav-Man-Wad98}.

The presence of the twisted boundary conditions on the bosonic field
makes it neccessary to redefine variables so that a convenient mode expansion is possible.
\be
\tilde x^i(\sigma + 2\pi(A-1),t) \equiv x^i_A(\sigma,t),
\sigma\in [0,2\pi )
\ee
$\tilde x^i$ has period  $2\pi Q_1Q_5R$.
It is easy to write the normal mode expansion 
\bea
\tilde x^i(\sigma,t) &=& (4\pi)^{-1/2} 
\sum_{n>0} [(
 \frac{a^i_n}{\sqrt n} e^{i n(-t+\sigma)/Q_1Q_5} 
\nn\\
&+& \frac{\tilde a^i_n}{\sqrt n} e^{i n(-t-\sigma)/Q_1Q_5} )
+ {\rm h.c.} ]\nn\\
\eea
The effect of the twisting on these oscillators is given by
\bea
\label{twist2}
g:a^i_n &\to& a^i_n e^{2\pi i n/Q_1Q_5} \nonumber\\
g:\tilde a^i_n &\to& \tilde a^i_n e^{-2\pi i n/Q_1Q_5}
\nonumber\\
\eea

The black hole state can now be explicitly constructed using the above oscillators,
\be
\label{eq-state}
|i\rangle = \prod_{n=1}^{\infty}\prod_{i} C(n,i) (a^{i\dagger}_n
)^{N_{L,n}^i} (\tilde a^{i\dagger}_n )^{N_{R,n}^i} |0 \rangle 
\ee
where $C(n,i)$ are normalization constants ensuring unit norm of the
state. $|0 \rangle$ represents the NS ground state. 

The present discussion is also valid in the Ramond sector, in which
case the ground state will have an additional spinor index but that
will not affect the $S$-matrix. This comment is important because the
black hole states are in the Ramond sector of the SCFT.This can be
inferred from the boundary conditions on the killing spinors in the
black hole background. For $AdS_3$ the dual boundary states are in the
Neveu-Schwarz sector of the SCFT \cite{henn}.

The creation operators create KK
(Kaluza-Klein) momentum along $S^1$
(parametrized by $x^5$). The total left (right) moving KK momentum of
\eq{eq-state} (in units of $1/\tilde R, \tilde R \equiv Q_1Q_5 R_5$,
$R_5$ being the radius of the $S^1$) is $N_L$ ($N_R$),
where
\be
\label{total-n}
N_L = \sum_{n,i} n N^i_{L,n}, \quad N_R = \sum_{n,i} n N^i_{R,n} 
\ee
The total KK momentum carried by the state $|i\rangle$ is an integer given by
\be
p_5= (N_L - N_R)/\tilde R
\ee 
where $\tilde R=Q_1 Q_5 R_5$. This fact also implies that $|i\rangle$
is invariant under the twisting action.

Now we can calculate the S-matrix element for the process:
\be
h_{89}(w,0) \to x^8_L(w/2,-w/2) + x^9_R((w/2,w/2)
\ee
(the numbers in parenthesis denote $(k_0, k_5)$)
\be
\label{s-matrix}
S_{if} = \frac{\sqrt{2} \kappa_5 w_1 w_2 \tilde R 
\delta_{n_1,n_2} 2\pi
\delta(w-w_1 - w_2)}{\sqrt{w_1 \tilde R
w_2 \tilde R w V_4}} \sqrt{N_{L,n_1}^8} 
\sqrt{\tilde N_{R,n_2}^9}
\ee
$V_4$ = volume of the noncompact space.
$N^i_{L,n}$ and $N^i_{R,n}$ denotes number
of oscillators with left- and right-moving
momentum $n$ respectively (see \eq{eq-state}). The
factors $\sqrt{N}$ appear from the normalization of the states
\be
|N\rangle = (a^\dagger)^N/\sqrt{N!} |0\rangle, [a, a^\dagger]=1
\ee
and the fact that 
\be
\langle N-1 | a |N \rangle = \sqrt{N}
\ee

>From the above S-matrix element we can evaluate the absorption probability for a quantum of frequency $w$
\bea
{\rm Prob}_{\rm abs}
&=& \frac{1}{\Omega _i}\sum_{i,f}| S_{if} |^2
\nn\\ 
&=&  \frac{\tilde R T}{V_4} \kappa_5^2 w 
\langle N_{iL}(w/2) \rangle\; 
\langle  N_{iR}(w/2) \rangle
\nn\\
\eea
Here $T$ is the total time of the process. 

The decay probability is obtained by the formula
\bea
{\rm Prob}_{\rm decay}
&=& \frac{1}{\Omega _f} \sum_{i,f}| S_{if} |^2
\nn\\ 
&=&  \frac{\tilde R T}{V_4} \kappa_5^2 w 
\langle N_{fL}(w/2) \rangle\; 
\langle N_{fR}(w/2) \rangle
\eea

We are interested in the process
\be
N_{iL,iR}(n_1) = N_{fL,fR}(n_1)+1
\ee
where $n_1/RQ_1Q_5=w/2$. To compare the string calculation 
with the semi-classical absorption calculation, 
where the black hole does not emit, we have to subtract the Prob$_{\rm
decay}$ from Prob$_{\rm abs}$.

A straight forward calculation then leads to 
\be
\label{abs-string}
\sigma_{abs} = 2 \pi^2 r_1^2 r_5^2 \frac{\pi w}{2} 
\frac{\exp(w/T_H)-1}{(\exp(w/2T_R)-1) (\exp(w/2T_L)-1)}
\ee
which exactly agrees with the semi-classical calculation (\ref{class-abs}).

Finally the decay rate is given by,
\be
\Gamma = \rm Prob_{\rm decay}\frac{V_4}{{\tilde R T}} 
\frac{d^4k}{(2 \pi)^4}
\ee
and hence,
\be
\label{decay-string}
\Gamma_H = \sigma_{abs} (e^{w/T_H}-1)^{-1} 
\frac{d^4k}{(2 \pi)^4}
\ee
which also exactly reproduces the semiclassical result (\ref{decay}).

We now make a few comments about these results:
\begin{itemize}

\item We have done a lot of work to be able to calculate the
absorption crossection and the Hawking rates which agree with the
semi-classical supergravity calculations. The string theory
calculations were originally done in \cite {Dha-Man-Wad96,Das-Mat96} 
and were based on a
model that was physically motivated by string dualities \cite {Mal-Sus96,Das-Mat}.
In particular the calculation in \cite{Das-Mat96} based on the DBI action
reproduced even the exact coefficient that matched with the
semi-classical answer for the absorption cross section of the minimal
scalars.  However this method did not work when applied to the fixed
scalars \cite {Cal-Gub-Kle96,Kra-Kle97,Kra-Kle96}.  This fact was very 
discouraging because it meant
the absence of a consistent starting point for string theory
calculations. The discovery of Maldacena \cite {malda-dual} finally enabled the
string theory calculations \cite {Dav-Man-Wad98, Dav-Man-Wad99} because it was able to make a precise
connection of the near horizon geometry with the infra-red fixed point
theory of brane dynamics.

\item The Hawking radiation calculation that we have presented is physically
motivated. However this method or twisted oscillators cannot be used
to calculate the rates corresponding to the particles whose vertex
operators come from the twisted sector. However these can also be done
using a formulation that relates the Hawking rates to the thermal
2-point function of the corresponding operators. Such a formulation
follows directly from the basic equations
(\ref{probdecay}),(\ref{probabs}). This approach has been discussed in
\cite{Dav-thesis} and it is originally due to Callan and Gubser \cite{gubserthesis}.

\item We recall that the semi-classical calculation was done in an asymptotically flat geometry and yet the absorption cross section matched with the SCFT calculation which is dual to the near horizon geometry. This emphasizes the fact that in the semi-classical calculation the absorption occurred entirely from 
near the horizon of the black hole.

\item It is important to point out that in the Maldacena limit the
closed string modes like the graviton decouple from the brane
system. This means that as $\alpha ' \rightarrow 0$ the interaction
hamiltonian of the graviton and the SCFT would be vanishingly
small. Hence one should not work in the strict $\alpha ' =0$
limit. However one can still obtain a sizable absorption cross
section. To see this it is sufficient to note (\ref{abs}), 
\be
\sigma_{abs}(w \rightarrow 0) = A_h = 4G_5\sqrt{Q_1Q_5N} 
\ee 
>From here we see that 
\be
A_h \sim \frac{g_6^2}{R_5} \alpha'^2 \sqrt{Q_1Q_5N} 
\ee
The quantity under the square root sign grows tends to infinity in the
supergravity limit and hence compensates the fact that $\alpha '
\rightarrow 0$.

\item As we have discussed before in section 9 non-renormalization
theorems garuntee the validity of the SCFT in the strong coupling
region. This also means that the 2-point functions of operators
belonging to the short multiplets, which determine the Hawking rates,
do not get renormalized. This is in particular true for all the 20
minimal scalars. Hence the Hawking rates of these particles are indeed
calculated and matched in the supergravity regime. Also note that the rates of all the 20 particles can be matched by fixing the normalization of any one of them, using the AdS/SCFT correspondence.

\item We have explained earlier the importance of studying the $D_1-D_5$
system in the presence of the vevs of $B_{NS}$. This corresponds to
stable rather than marginal bound states and non-singular SCFT. This raises the question, whether the Hawking rates depend upon these vevs. They do not. This was shown in \cite{Dav-Man-Wad99}.
\end{itemize}

\section{Concluding Remarks}
Let us conclude by stating some of the outstanding problems.
\begin{enumerate}

\item How does one formulate the effective long wave length theory of the 
non-supersymmetric black holes? 

\item How does one derive space-time from brane theory? In particular
is there a way of deducing $AdS_3\times S^3$ (the infinitely stretched
horizon) as a consequence of brane dynamics? 
The method of coadjoint orbits is a promising approach to this question.
And what about the black
hole horizon itself. These questions are intimately tied to explaining the
geometric Bekenstein-Hawking formula or in other words understand the holographic principle \cite{tHooft}. 

\item The $D_1-D_5$ system has relevant perturbations. It would be interesting to study the holographic renormalization group in this situation. What is the end point of the RG flow?

\end{enumerate}

\gap

{\large {\bf Acknowledgements}}

\pgap

It is a pleasure to acknowledge Justin David, Avinash Dhar, Fawad
Hassan and Gautam Mandal for useful discussions and collaboration on
most of the topics discussed in these notes. 

I thank Jose Labastida and Jose Barbon for inviting me to lecture at
the Advanced School in Santiago de Compostela, Debasish Ghoshal and
Dileep Jatkar for lectures given at the MRI workshop and Farhad
Ardalan and Hessam Arfei for the course given at the Isfahan String
School and Workshop. I thank all of them for their warm and spirited
hospitality.

I would also like to acknowledge discussions on the topics discussed
in these notes with Daniele Amati, Thibault Damour, Sumit Das, Juan
Maldacena, Kumar Shiv Narain, Ashoke Sen, Lenny Susskind, Gabriele
Veneziano and Edward Witten.

\end{document}